\documentclass[10pt,twoside]{article}
\usepackage{a4wide,amssymb,tildes,epsfig}
\usepackage[tight]{subfigure}
\usepackage{amsmath,fancyhdr}

\title{\textbf{Cellular buckling from mode interaction in I-beams
    under uniform bending}}

\author{\normalsize\textsc{By M. Ahmer Wadee}\footnote{Author for
    correspondence: a.wadee@imperial.ac.uk}, \textsc{and Leroy Gardner}\\~\\
  \normalsize \emph{Department of Civil \& Environmental
    Engineering,}\\ \normalsize\emph{Imperial College London, 
      London SW7 2AZ, UK}}

\date{}

\begin{document}
\maketitle

\thispagestyle{fancy}
\renewcommand{\headrulewidth}{0pt}
\renewcommand\footnoterule{}

\newcommand{\Dz}{\mbox{$\mathrm{d}z$}}
\newcommand{\Mcr}{\mbox{$M_\mathrm{cr}$}}

\begin{abstract}
  Beams made from thin-walled elements, whilst very efficient in terms
  of the structural strength and stiffness to weight ratios, can be
  susceptible to highly complex instability phenomena.  A nonlinear
  analytical formulation based on variational principles for the
  ubiquitous I-beam with thin flanges under uniform bending is
  presented. The resulting system of differential and integral
  equations are solved using numerical continuation techniques such
  that the response far into the post-buckling range can be
  portrayed. The interaction between global lateral-torsional buckling
  of the beam and local buckling of the flange plate is found to
  oblige the buckling deformation to localize initially at the beam
  midspan with subsequent cellular buckling (snaking) being predicted
  theoretically for the first time. Solutions from the model compare
  very favourably with a series of classic experiments and some newly
  conducted tests which also exhibit the predicted sequence of
  localized followed by cellular buckling.
\end{abstract}

\section{Introduction}
\label{sec:intro}

Beams are possibly the most common type of structural component, but
when made from thin metallic plate elements they are well known to
suffer from a variety of elastic instability phenomena.  There has
been a vast amount of research into the buckling of thin-walled
structural components with much insight gained
\cite{Hancock78,Schafer2002,CIMS2008_vol2}. In the current work, the
well-known problem of a beam made from a linear elastic material with
an open and doubly-symmetric cross-section -- an ``I-beam'' -- under
uniform bending about the strong axis is studied in detail.  Under
this type of loading, long beams are primarily susceptible to a global
mode of instability namely lateral-torsional buckling (LTB), where, as
the name suggests, the beam deflects sideways and twists once a
threshold critical moment is reached \cite{TG61}. However, when the
individual plate elements of the beam cross-section, namely the
flanges and the web, are relatively thin or slender, elastic local
buckling of these may also occur; if this happens in combination with
global instability, then the resulting behaviour is usually far more
unstable than when the modes are triggered individually. Other
structural components, usually those under axial compression rather
than uniform bending, such as I-section struts with thin plate flanges
\cite{Becque_Rasmussen_expt_ASCE2009}, sandwich struts
\cite{MAW_prs98}, stringer-stiffened and corrugated plates
\cite{Koiter76,Pignataro_TWS2000} and built-up, compound or
reticulated columns \cite{TH73} are all known to suffer from so-called
\emph{interactive buckling} phenomena where the global and local modes
of instability combine nonlinearly.

Experimental evidence suggests that the destabilization from the mode
interaction of LTB and flange local buckling is severe
\cite{cherry_ltb_local_1960,MGS91}, the response being highly
sensitive to geometric imperfections particularly when the critical
loads coincide
\cite{Mollmann1_ijss_89,Mollmann2_ijss_89}. Nevertheless, apart from
the aforementioned work where some successful numerical modelling
(both finite strip and finite element) and qualitative modelling using
rigid links and spring elements were presented, the formulation of a
mathematical model accounting for the interactive behaviour has not
been forthcoming. The current work presents the development of a
variational model that accounts for the mode interaction between
global LTB and local buckling of a flange such that the perfect
elastic post-buckling response of the beam can be evaluated. A system
of nonlinear ordinary differential equations subject to integral
constraints is derived, which are solved using the numerical
continuation package \textsc{Auto} \cite{Auto2007}. It is indeed found
that the system is highly unstable when interactive buckling is
triggered; snap-backs in the response showing sequential
destabilization and restabilization and a progressive spreading of the
initial localized buckling mode are also revealed. This latter type of
response has become known in the literature as \emph{cellular
  buckling} \cite{nondyn} or \emph{snaking} \cite{BurkeKnobloch2007}
and it is shown to appear naturally in the numerical results of the
current model. As far as the authors are aware, this is the first time
this phenomenon has been found in beams undergoing LTB and local
buckling simultaneously. Similar behaviour has been discovered in
various other mechanical systems such as in the post-buckling of
cylindrical shells \cite{HLC98,HuntLordPeletier03} and the sequential
folding of geological layers \cite{MAW_jsg,MAW_jmps05}.

Experimental results generated for the current study and from the
literature are compared with the results from the presented model;
highly encouraging quantitative results emerge both in terms of the
mechanical destabilization exhibited and the nature of the
post-buckling deformation with tangible evidence of cellular buckling
being found. This demonstrates that the fundamental physics of this
system is captured by the analytical approach. A brief discussion is
presented on whether other structural components made from thin-walled
elements may also exhibit cellular buckling when local and global
modes of instability interact. Conclusions are then drawn.

\section{Variational formulation}

\subsection{Classical critical moment derivation via energy}
\label{sec:Mcr}

\begin{figure}[htb]
  \centering
  \subfigure[Beam geometry: elevation (left) and cross-section (right)]{%
    \psfig{figure=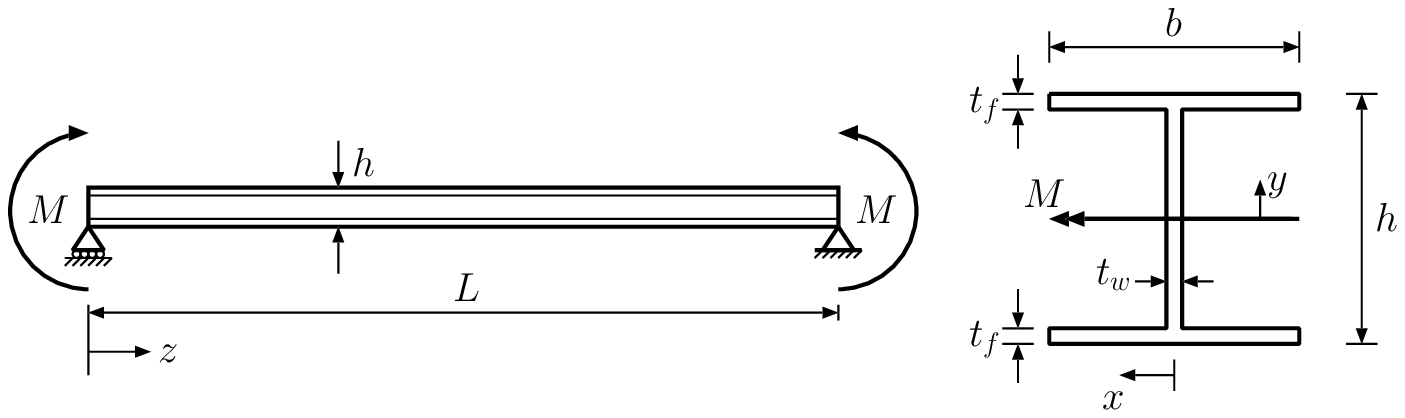,width=95mm}}\quad
  \subfigure[LTB displacements]{\psfig{figure=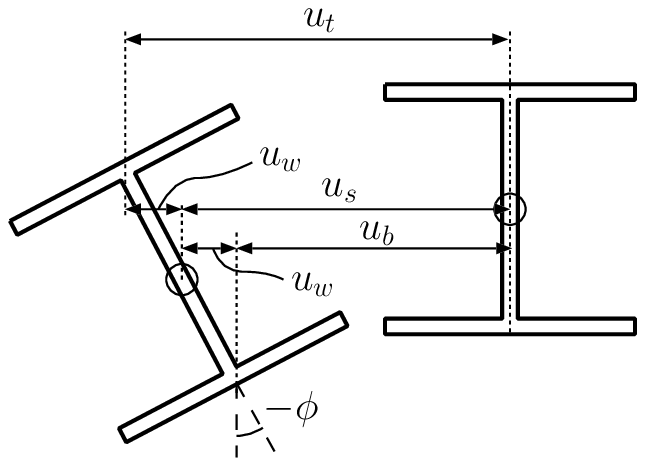,width=45mm}}
  \caption{(a) An I-section beam under uniform strong axis moment
    $M$. (b) Lateral displacements and torsional rotations due to
    LTB. A right-hand Cartesian coordinate system is employed.}
  \label{fig:Isection}
\end{figure}

\begin{figure}[htb]
  \centering
  \subfigure[Axial stresses due to bending]{%
    \psfig{figure=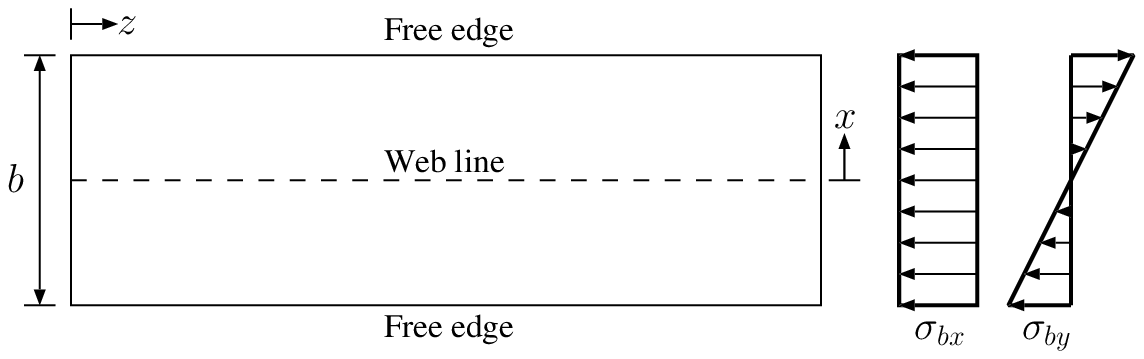,height=30mm}}\quad
  \subfigure[Bending moments under LTB]{\psfig{figure=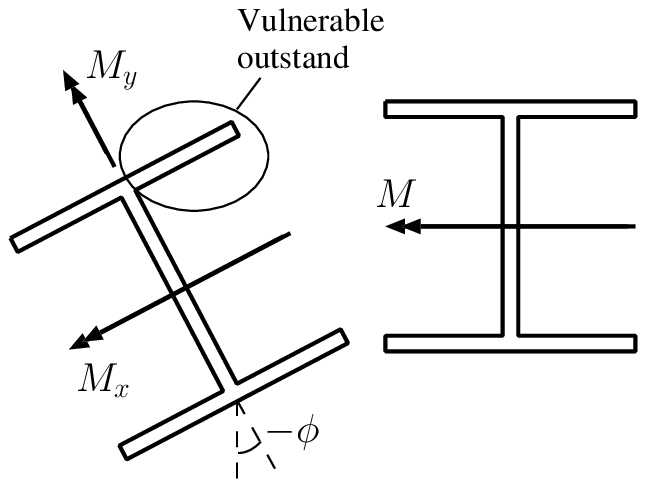,width=45mm}}
  \caption{After LTB: (a) top flange stress components from strong
    ($\sigma_{bx}$) and weak axis ($\sigma_{by}$) bending moments; (b)
    induced bending moment components during LTB, $M_x$ and $M_y$,
    which force one of the flange outstands to become more heavily
    compressed and hence vulnerable to local buckling.}
  \label{fig:topflange}
\end{figure}

Consider a uniform simply-supported (pinned) doubly-symmetric I-beam
of length $L$ made from an isotropic and homogeneous linear elastic
material with Young's modulus $E$ and Poisson's ratio $\nu$. The
overall beam height and flange widths are $h$ and $b$ respectively
with the web and flange thicknesses being $t_w$ and $t_f$
respectively. The beam is under bending about the strong axis with a
uniform moment $M$, as shown in Figure \ref{fig:Isection}(a).
The second moments of area about the strong and weak axes are defined
as $I_x$ and $I_y$ respectively. When LTB occurs, as noted in the
literature \cite{TG61}, the strong axis bending moment and
corresponding displacements only couple with the lateral displacements
and torsional rotations at higher orders. For this reason and for the
sake of simplicity, the displacements arising from strong axis bending
are presently neglected, which is a common assumption; in future work,
however, these coupling effects may be incorporated. Therefore, only
two separate lateral displacements are defined for determining the
respective positions of the local centroids of the web and the
flanges: $u_s$ and $u_w$.  The lateral displacements of the top
($u_t$) and bottom flanges ($u_b$) respectively are thus (see Figure
\ref{fig:Isection}(b)):
\begin{equation}
  u_t = u_s + u_w, \quad
  u_b = u_s - u_w.
\end{equation}
Moreover, a torsional angle of magnitude $\phi$ to the vertical is
introduced and the applied moment $M$ is transformed thus into components
of strong and weak axis moments, $M_x$ and $M_y$ respectively:
\begin{equation}
  M_x = M \cos (-\phi) \approx M, \quad
  M_y = M \sin (-\phi) \approx -M \phi,
\end{equation}
both expressions being written to the leading order. Figure
\ref{fig:topflange}(a)
shows a plan view of the top flange of the beam and the stress
distribution components from the strong axis and the induced weak axis
bending moments.  Therefore, the maximum strong axis compressive
stress occurs when $y=h/2$ and coupling this with the co-existing
compressive component of the weak axis stress, the most vulnerable
flange outstand is defined, as shown in Figure
\ref{fig:topflange}(b). Since the important component of bending under
LTB is about the weak axis, the values of the relevant second moment
of area for the flange ($I_f$) can be approximated to $I_y/2$ of the
whole section, where $I_y=b^3 t_f/6$, by assuming that the overall
contribution of the web is very small, which is true for I-beams of
practical dimensions. The strain energy stored in the beam due to
bending ($U_b$) is therefore:
\begin{equation}
  U_b = \frac12 EI_f \int_0^L \left( u_t''^2 + u_b''^2 \right) \D z =
  \frac12 EI_y \int_0^L u_s''^2 \D z + \frac12 EI_w \int_0^L \phi''^2 \D z,
\end{equation}
where primes denote differentiation with respect to the axial
coordinate $z$, $u_w=h\phi/2$ and $I_w=I_y h^2/4$. The strain energy
stored from uniform torsion ($U_T$) is:
\begin{equation}
  U_T = \frac12 \int_0^L T \D \phi = \frac12 GJ \int_0^L \phi'^2 \D z,
\end{equation}
where $T$ is the torque, $G$ is the material shear modulus and $J$ is
the St Venant torsion constant defined for an I-section as:
\begin{equation}
  \label{eq:J}
  J = \frac13 \left[ 2 bt_f^3 + (h-2t_f)t_w^3 \right].
\end{equation}
The work done by the external moment $M\Theta$ is given by the induced
weak axis moment $M_y$ multiplied by the average end rotation from
bending about the weak axis; this is given by the following expression
\cite{Pi_Trahair_Rajasekaran_1992}:
\begin{equation}
  M\Theta = \int_0^L M \phi u_s'' \D z
\end{equation}
and so the total potential energy is thus:
\begin{equation}
  V = U_b + U_T - M\Theta = \int_0^L \left[ \frac12 EI_y u_s''^2 +
    \frac12 EI_w \phi''^2 + \frac12 GJ \phi'^2 - M \phi u_s'' \right] \Dx z.
  \label{eq:V_LTB}
\end{equation}

To find the critical moment $\Mcr$, the calculus of variations could
be used to derive the governing differential equation. However, since
the solution of the buckling mode is known to be sinusoidal
\cite{TG61}, the same result can be achieved by using a two degree of
freedom Rayleigh--Ritz formulation with the following trial functions:
\begin{equation}
  u_s = q_s L \sin \frac{\pi z}{L}, \quad
  \phi = -q_{\phi} \sin \frac{\pi z}{L},
\label{eq:dofs}
\end{equation}
where $q_s$ and $q_\phi$ are generalized coordinates defining the
amplitudes of $u_s$ and $\phi$ respectively. Note that owing to the
coordinate system used, the negative sign in front of $q_\phi$ is
necessary to ensure that $q_\phi>0$ when $q_s>0$.  Substituting the
expressions for $u_s$ and $\phi$ from Equation (\ref{eq:dofs}) gives:
\begin{equation}
  V = \int_0^L \left[ \frac{EI_y \pi^4}{2L^2} \sin^2 \frac{\pi z}{L} \left(
    q_s^2 + q_\phi^2 \lambda^2 \right) + \frac{GJ \pi^2}{2L^2}
  q_\phi^2 \cos^2 \frac{\pi z}{L} - \frac{M}{L} q_s q_\phi
  \pi^2 \sin^2 \frac{\pi z}{L} \right] \Dx z
\end{equation}
where $\lambda=h/2L$.  The formulation is in small deflections
(linear) and so only a critical equilibrium analysis is possible at
this stage. The advantage of using the Rayleigh--Ritz method becomes
more apparent in the interactive buckling model below. Performing the
integration and then assembling the Hessian matrix $\mathbf{V}_{ij}$
at the critical point $\mathrm{C}$, gives the following condition:
\begin{equation}
  \mathbf{V}_{ij}^\mathrm{C} 
  = \left|
    \begin{array}{cc}
      V^\mathrm{C}_{ss} & V^\mathrm{C}_{s\phi} \\
      V^\mathrm{C}_{\phi s} & V^\mathrm{C}_{\phi \phi}
    \end{array} \right|
  = \left|
    \begin{array}{cc}
      EI_y (\pi/L)^4 & - \Mcr (\pi/L)^2 \\
      - \Mcr (\pi/L)^2 & GJ (\pi/L)^2 + EI_w (\pi/L)^4\\
    \end{array} \right| = 0,
  \label{eq:VijC}
\end{equation}
where the individual elements of the matrix
$\mathbf{V}_{ij}^\mathrm{C}$ are thus:
\begin{equation}
  \label{eq:Vab}
  V_{ab}^\mathrm{C} = \left. \frac{\partial^2 V}{\partial q_a \partial
      q_b} \right|^\mathrm{C}.
\end{equation}
Solving equation (\ref{eq:VijC}) gives the classical expression for
the critical moment $\Mcr$ that triggers LTB for a beam with a
doubly-symmetric cross-section under uniform bending \cite{TG61}:
\begin{equation}
  \label{eq:Mcr}
  \Mcr = \frac{\pi}{L}\sqrt{EI_y GJ}\sqrt{1 +
      \frac{\pi^2}{L^2} \frac{EI_w}{GJ}}.
\end{equation}

\subsection{Interactive buckling model}
\label{sec:Inter}

From the previous section, it has been shown that as the displacements
and rotations from LTB grow the applied moment $M$ can be expressed as
a component about the strong axis ($M_x$) and an induced component
about the weak axis ($M_y$) at any point along $z$. As a result, the
vulnerable outstand of the flange, as identified in Figure
\ref{fig:topflange}, may therefore buckle locally as a plate.  The
critical stress of plate buckling for a uniaxially and uniformly
compressed rectangular plate, with one long edge pinned and the other
free, is given by the well known formula \cite{TG61}:
\begin{equation}
  \label{eq:local_sigma}
  \sigma^\mathrm{C}_\mathrm{Local} = 0.43 \frac{\pi^2 D}{b_f^2 t_f},
\end{equation}
where, in the current case, $b_f$ is the width of the vulnerable
flange outstand and is given by $(b-t_w)/2$ and $D$ is the flexural
rigidity of the flange plate that is equal to
$Et_f^3/[12(1-\nu^2)]$. This addresses the case for the flange
buckling locally before any LTB occurs.

\begin{figure}[htb]
  \centering
  \subfigure[Top flange]{\psfig{figure=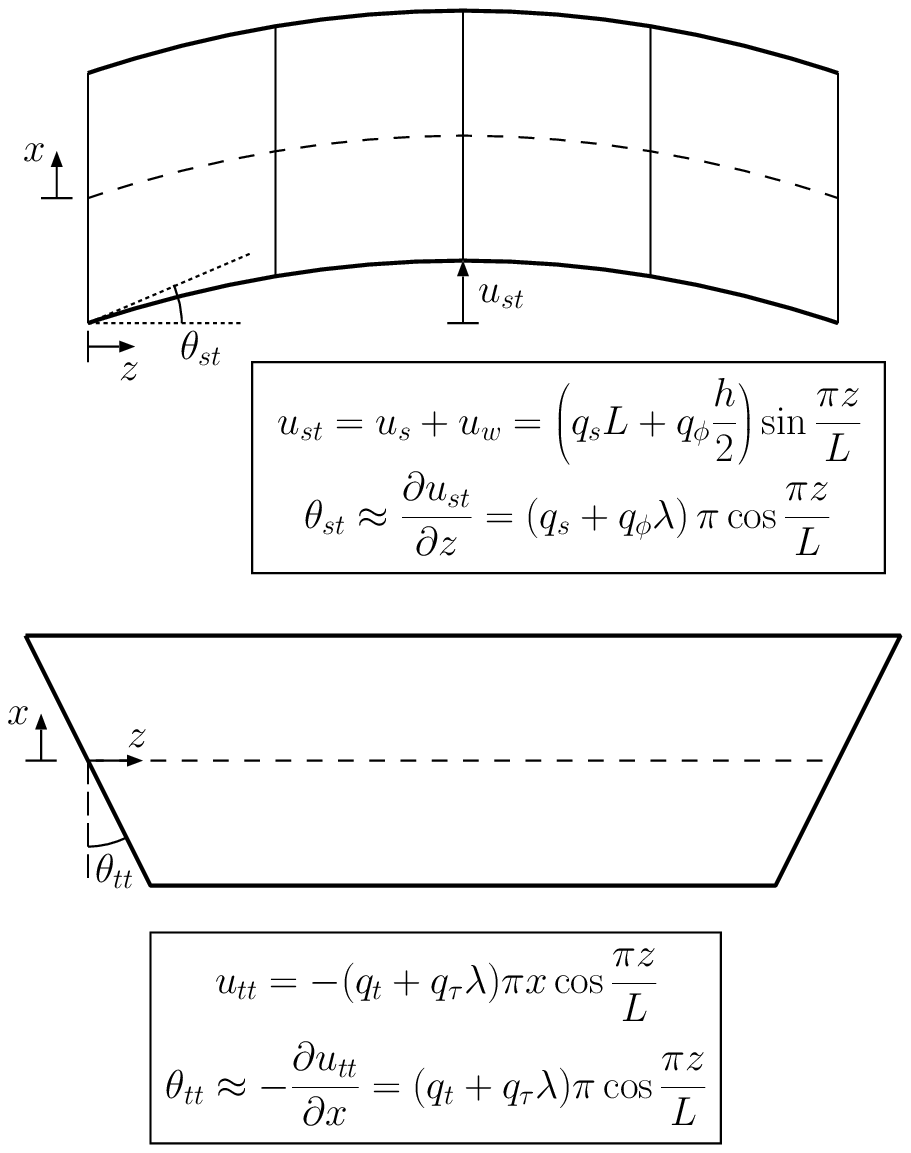,width=65mm}}
  \subfigure[Bottom flange]{\psfig{figure=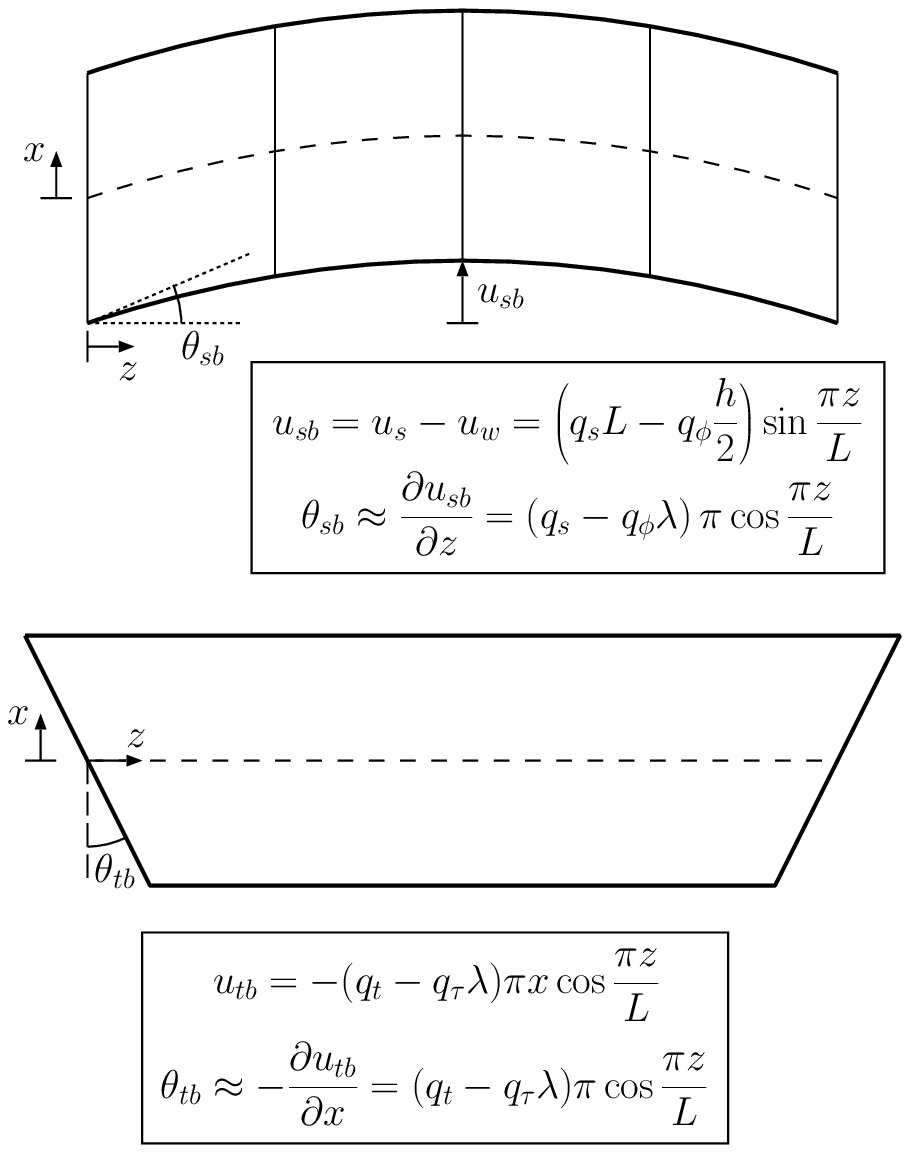,width=65mm}}
  \caption{Sway (upper row) and tilt (lower row) modes of LTB in both
    flanges.}
  \label{fig:swaytilt_top_bottom}
\end{figure}

It was shown in \cite{MAW_prs98} that the intrinsic assumptions in
Euler--Bernoulli bending theory were insufficient to model any
interaction between global and local buckling modes. The allowance of
shear strains to develop within the individual elements, however
small, being key to the formulation. Figure
\ref{fig:swaytilt_top_bottom}
shows a useful way that shear can be introduced; the displacement of
the top and bottom flanges being decomposed into separate ``sway'' and
``tilt'' modes \cite{HDSM88} after the global instability (LTB) has
been triggered. Each original generalized coordinate $q_s$ and
$q_\phi$ is defined as a ``sway'' and has a corresponding ``tilt''
component, with associated generalized coordinates $q_t$ and $q_\tau$
respectively. This is akin to a Timoshenko beam formulation
\cite{Shear}; when $q_s \neq q_t$ and $q_\phi \neq q_\tau$, shear
strains are developed and allow the potential for modelling
simultaneous LTB and local buckling.

\begin{figure}[htb]
  \centering
  \subfigure[$u(x,z)$]{\psfig{figure=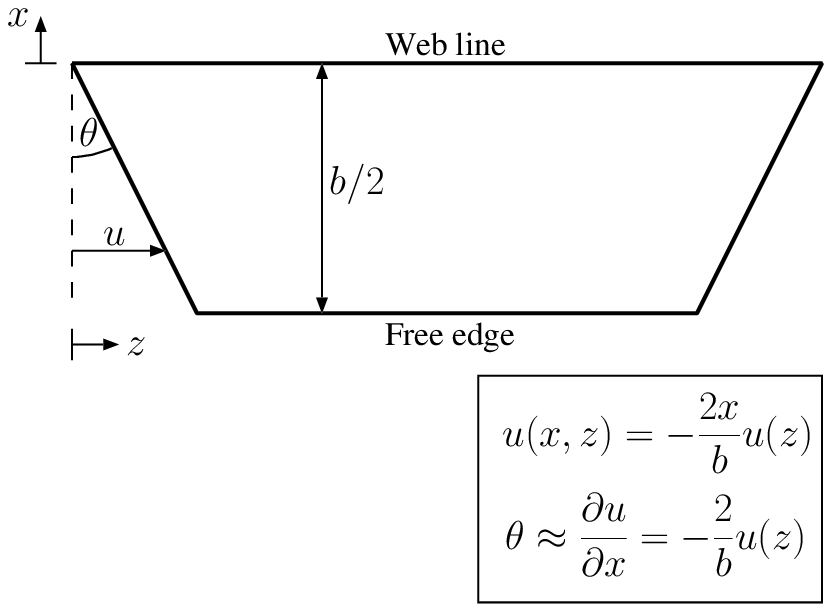,width=70mm}}
  \subfigure[$w(x,z)$]{\psfig{figure=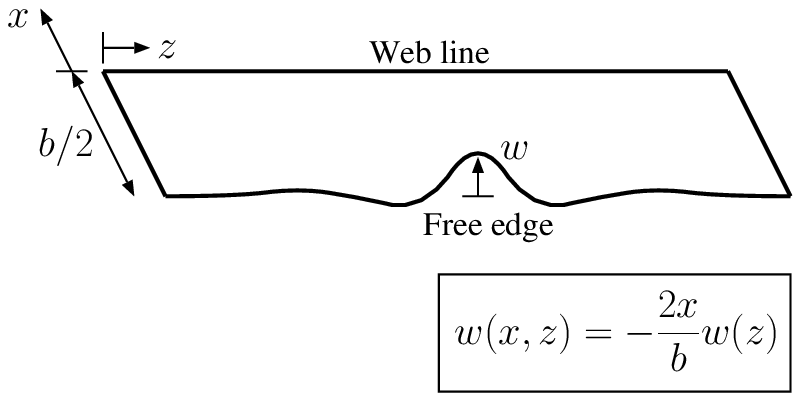,width=70mm}}
  \caption{Displacement functions $u$ and $w$ defined for the
    vulnerable part of the top flange: $x=[-b/2,0]$ and
    $y=[h/2-t_f,h/2]$.}
  \label{fig:uwdef}
\end{figure}

Previous work on this type of interactive buckling has included
experimental work combined with effective width theory
\cite{cherry_ltb_local_1960}, some phenomenological modelling using
rigid links and springs along with experiments \cite{MGS91}, some
numerical work using a finite strip formulation
\cite{Mollmann1_ijss_89,Mollmann2_ijss_89} and a finite element
formulation \cite{MSGP97}. To model this analytically, however, two
displacement functions to account for the extra in-plane displacement
$u$ and out-of-plane displacement $w$ (Figure \ref{fig:uwdef})
need to be defined. Since the strain from the weak axis moment is
linear in $x$ and that the boundary condition for the line of the web
is pinned out of the plane of the flange, the following linear
distribution in $x$ can be assumed for $u$ and $w$:
\begin{equation}
  u(x,z) = -\frac{2x}{b} u(z), \quad
  w(x,z) = -\frac{2x}{b} w(z).
\end{equation}
It is worth noting that restricting both interactive buckling
displacement functions $u$ and $w$ to the vulnerable part of the
compression flange confines the current model to cases where LTB
occurs before local buckling. In the cases where local buckling occurs
first, the system would be expected to trigger the global mode rapidly
afterwards \cite{MAW_ijss00} and the current model can be used to
indicate the deformation levels where the interaction would occur (as
seen later). However, to obtain an accurate linear eigenvalue solution
for pure local buckling in the current framework, at least another set
of in-plane and out-of-plane displacement functions, replicating the
role of $u$ and $w$ respectively, would need to be defined for the
non-vulnerable part of the compression flange; this addition is left
for future work.

\subsubsection{Local bending energy}

Experimental evidence, presented in \cite{MSGP97}, suggests that
during interactive buckling the significant local out-of-plane
displacements within the flanges are confined to the vulnerable
outstand. The component of additional strain energy stored in bending
$U_{bl}$ is hence given by:
\begin{equation}
  U_{bl} = \frac{D}{2} \int_0^L \int_{-b/2}^0 \left\{
    \left( \frac{\partial^2 w}{\partial x^2} + \frac{\partial^2
        w}{\partial z^2} \right)^2 - 2(1-\nu) 
    \left[ 
      \frac{\partial^2 w}{\partial x^2} \frac{\partial^2 w}{\partial z^2}
      - \left(\frac{\partial^2 w}{\partial x \partial z}\right)^2
    \right] \right\} \,\mathrm{d}x \,\mathrm{d}z.
\end{equation}
Substituting $w$ into $U_{bl}$, the following expression is obtained:
\begin{equation}
  U_{bl} = \frac{D}{2 b} \int_0^L \left[
      \frac{b^2}{6} w''^2 + 4 (1-\nu) w'^2 
    \right] \, \mathrm{d}z.
\end{equation}

\subsubsection{Flange energy from axial and shear strains}

Since the flanges are assumed to behave in the manner of Timoshenko
beams, the bending of the flanges when LTB occurs introduce both axial
and shear strains, $\varepsilon$ and $\gamma$ respectively.
The vulnerable part of the top flange, where $x=[-b/2,0]$ and
$y=[h/2-t_f,h/2]$, also has the possibility of local buckling; von
K\'arm\'an plate theory gives a standard expression for the axial
strain $\varepsilon$ in the $z$-direction that accounts for both LTB
and local buckling terms, thus:
\begin{equation}
    \varepsilon_{t1} = 
    \frac{\partial u_{tt}}{\partial z} + \frac{\partial u}{\partial z} +
    \frac12 \left( \frac{\partial w}{\partial z} \right)^2
    =  (q_t + q_\tau \lambda) \frac{\pi^2 x}{L} \sin \frac{\pi z}{L} -
    \frac{2x}{b} u' + \frac{2x^2}{b^2} w'^2.  
\end{equation}
For the part of the top flange that is not vulnerable to local
buckling, where $x=[0,b/2]$ and $y=[h/2-t_f,h/2]$, and the bottom
flange, where $y=[-h/2,-h/2+t_f]$, the following respective axial strain
expressions are obtained:
\begin{align}
    \varepsilon_{t2} &= 
    \frac{\partial u_{tt}}{\partial z} =  (q_t + q_\tau \lambda)
    \frac{\pi^2 x}{L} \sin \frac{\pi z}{L},\\
    \varepsilon_{b} &= 
    \frac{\partial u_{tb}}{\partial z}
    = (q_t - q_\tau \lambda)\frac{\pi^2 x}{L} \sin \frac{\pi z}{L}.
\end{align}
The standard strain energy expression is then integrated over the
volume of the flanges:
\begin{equation}
  \begin{split}
    U_m & = \frac{E}{2} \int_0^L \left[ \int_{h/2-t_f}^{h/2} \int_{-b/2}^0
      \varepsilon_{t1}^2 \D x \D y + \int_{h/2-t_f}^{h/2}
      \int_{0}^{b/2} \varepsilon_{t2}^2 \D x \D y +
      \int_{-h/2}^{-h/2+t_f} \int_{-b/2}^{b/2} \varepsilon_{b}^2 \D x
      \D y \right] \Dx z\\
    &= \frac{Ebt_f}{12} \int_0^L \biggl[ u'^2 + \frac{3}{20} w'^4 +
    \frac{3}{4} u'w'^2 - (q_t + q_\tau \lambda) \pi^2 \psi \sin
    \frac{\pi z}{L}
    \left( u' + \frac38 w'^2 \right) \\
    & \qquad \qquad \qquad + \left( q_t^2 + q_\tau^2 \lambda^2 \right)
    \pi^4 \psi^2 \sin^2 \frac{\pi z}{L} \biggr] \D z,
  \end{split}
\end{equation}
where $\psi=b/L$ is a beam aspect ratio parameter. It is worth noting
that the transverse displacement and the strain in the $x$-direction
are omitted from the current formulation. This is a simplification
derived from \cite{Koiter76}, where these components were found to
have a negligible effect on the post-buckling stiffness of a
uniaxially compressed long plate with one longitudinal edge being
pinned and the other being free.

In terms of the shear strain $\gamma$ in the $xz$ plane within the top
flange, where $y=[h/2-t_f,h/2]$, von K\'arm\'an plate theory gives a
standard expression, which needs to account for both LTB and local
terms for the vulnerable outstand $x=[-b/2,0]$:
\begin{equation}
  \begin{split}
    \gamma_{t1} & = \theta_{st} - \theta_{tt} + \text{local buckling terms}\\
    &= \frac{\partial u_{st}}{\partial z} + \frac{\partial u_{tt}}{\partial x}
    + \frac{\partial u}{\partial x} + \frac{\partial w}{\partial
      x}\frac{\partial w}{\partial z}  
    = \left[(q_s-q_t) + (q_\phi - q_\tau )\lambda \right] 
    \pi \cos \frac{\pi z}{L}- \frac{2}{b} u + \frac{4 x}{b^2} w w',
  \end{split}
\end{equation}
and purely LTB terms for the non-vulnerable part of the top flange,
$x=[0,b/2]$, and the bottom flange, where $y=[-h/2,-h/2+t_f]$,
respectively:
\begin{align}
    \gamma_{t2} &= \theta_{st} - \theta_{tt}
    = \frac{\partial u_{st}}{\partial z} + \frac{\partial u_{tt}}{\partial x}
    = \left[(q_s-q_t) + (q_\phi - q_\tau )\lambda \right] 
    \pi \cos \frac{\pi z}{L},\\
    \gamma_{b} &= \theta_{sb} - \theta_{tb}
    = \frac{\partial u_{sb}}{\partial z} + \frac{\partial u_{tb}}{\partial x}
    = \left[(q_s-q_t) - (q_\phi - q_\tau )\lambda \right] 
    \pi \cos \frac{\pi z}{L}.
\end{align}
After obtaining the shear strains, the standard strain energy
expression needs to be integrated over the volume of the flanges,
thus:
\begin{equation}
  \begin{split}
    U_s &= \frac{G}{2} \int_0^L \left[ \int_{h/2-t_f}^{h/2}
      \int_{-b/2}^0 \gamma_{t1}^2 \D x \D y + 
    \int_{h/2-t_f}^{h/2} \int_0^{b/2} \gamma_{t2}^2 \D x \D y
    + \int_{-h/2}^{-h/2+t_f} \int_{-b/2}^{b/2} \gamma_{b}^2
    \D x \D y \right] \Dx z\\
    &= \frac{Gbt_f}{2} \int_0^L \biggl\{
      \frac{2}{b^2} \left( u^2 + \frac13 w^2 w'^2 + uww' \right)
      - \left[ (q_s - q_t) + (q_\phi - q_\tau) \lambda \right] \frac{\pi}{b}
      \cos \frac{\pi z}{L} \left(2u + ww' \right) \\
    & \qquad \qquad \qquad
      + \left[ (q_s - q_t)^2 + (q_\phi - q_\tau)^2 \lambda^2 \right]
      2\pi^2 \cos^2 \frac{\pi z}{L} \biggr\} \D z.
  \end{split}
\end{equation}

\subsubsection{Work done contribution}

\begin{figure}[htb]
  \centerline{\psfig{figure=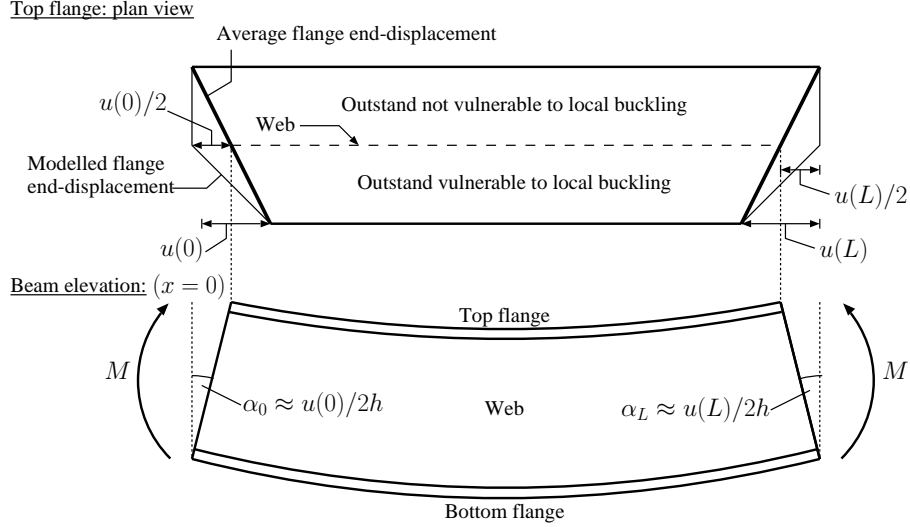,width=120mm}}
  \caption{Contribution from local buckling in the work done term,
    where $h$ is the overall cross-section height.}
  \label{fig:localwd}
\end{figure}

An additional contribution from the vulnerable flange's local in-plane
displacement function $u$ needs to be included in the work done
term. Figure \ref{fig:localwd}
shows that the compression flange has a distribution of in-plane
displacement which is assumed to have an average linear distribution
in $x$. Including this as an average end-rotation, angles $\alpha_0$
and $\alpha_L$ are obtained as shown in the diagram. Since the end
rotation angles can be expressed in terms of the local in-plane
displacement function $u$, the expression for the local contribution
to the work done is:
\begin{equation}
  \label{eq:localwd}
  M\Theta_l = M_x (\alpha_0 - \alpha_L) = - \frac{M}{2h} \int_0^L u' \D z. 
\end{equation}

\subsubsection{Total potential energy}

The expression for the total potential energy of the interactive
buckling model system can be written as a sum of the individual terms
of the strain energies minus the work done terms from \S\ref{sec:Mcr}
and \S\ref{sec:Inter}, with $U_m$ and $U_s$ replacing $U_b$ to account
for the change in the bending theory assumptions:
\begin{equation}
  \label{eq:TPE}
  \begin{split}
    V = U_m + U_T + U_{bl} + U_s - M(\Theta + \Theta_l).
  \end{split}  
\end{equation}
This new energy function $V$ replaces equation (\ref{eq:V_LTB}) and is
written in terms of non-dimensionalized variables that replace the
original ones, thus:
\begin{equation}
  \label{eq:nondim}
  \tilde{u} = \frac{2}{L}u, \quad
  \tilde{w} = \frac{2}{L}w, \quad
  \tilde{z} = \frac{2}{L}z,
\end{equation}
giving the full expression for $V$, where primes henceforth represent
derivatives with respect to $\tilde{z}$:
\begin{equation}
  \label{eq:TPE_nondim}
  \begin{split}
    V = \frac{L}{2} \int_0^2 & \biggr\{ 
    \frac{Ebt_f}{12} \biggl[
      \tu'^2 + \frac{3}{20} \tw'^4 + \frac34 \tu' \tw'^2
      - (q_t + q_\tau \lambda) \pi^2 \psi \sin \frac{\pi \tz}{2}
      \left( \tu' + \frac38 \tw'^2 \right) \\ 
      &
      + \left( q_t^2 + q_\tau^2 \lambda^2
      \right) \pi^4 \psi^2 \sin^2 \frac{\pi \tz}{2} \biggr] + \frac{GJ
        \pi^2}{2L^2} q_\phi^2
    \cos^2 \frac{\pi \tz}{2} + \frac{D}{b} \left[ \frac{\psi^2}{3} \tw''^2
      + 2 (1-\nu) \tw'^2 \right] \\
    & 
     + \frac{Gbt_f}{2} \biggl[
      \frac{1}{2\psi^2} \left( u^2 + \frac13 \tw^2 \tw'^2 + \tu \tw \tw' \right)
      + \left[ (q_s - q_t)^2 + ( q_\phi - q_\tau )^2 \lambda^2 \right] 2\pi^2
      \cos^2 \frac{\pi \tz}{2} \\
      & 
      - \left[ (q_s - q_t) + (q_\phi -q_\tau ) \lambda \right] \frac{\pi}{\psi} 
      \cos \frac{\pi \tz}{2} \left(\tu + \frac12 \tw \tw' \right) \biggr]
      - \frac{M}{L} \left( q_s q_\phi \pi^2 \sin^2 \frac{\pi \tz}{2} -
        \frac{1}{4\lambda} \tu' \right) \biggr\} \Dx \tz.
  \end{split}
\end{equation}

\subsubsection{Linear eigenvalue analysis}

With $u$ and $w$ being zero, along with their derivatives, before any
local buckling occurs, the Hessian matrix $\mathbf{V}_{ij}$, now
including terms associated with the ``tilt'' generalized coordinates,
can still be used to find $\Mcr$ the critical moment for LTB. The
Hessian matrix is thus:
\begin{equation}
  \mathbf{V}_{ij}^\mathrm{F} = \left(
    \begin{array}{cccc}
      V_{ss} & V_{st} & V_{s\phi} & V_{s\tau}\\
      V_{ts} & V_{tt} & V_{t\phi} & V_{t\tau}\\
      V_{\phi s} & V_{\phi t} & V_{\phi\phi} & V_{\phi\tau}\\
      V_{\tau s} & V_{\tau t} & V_{\tau \phi} & V_{\tau \tau}
    \end{array} \right),
\end{equation}
with the individual terms being:
\begin{equation}
  \begin{split}
    V_{ss} &= GLbt_f\pi^2, \quad
    V_{tt} = \frac{EI_y\pi^4 \psi^2}{2L} + GLbt_f\pi^2, \quad
    V_{\phi\phi} = \frac{GJ\pi^2}{2L} + GLbt_f\pi^2,\\
    V_{\tau\tau} &= \frac{EI_w \pi^4}{2L^3} + GLbt_f\pi^2 \lambda^2 =
    \pi^2 \lambda^2 \left( \frac{EI_y \pi^2}{2L} + GLbt_f \right), \quad
    V_{st} = V_{ts} = - GLbt_f \pi^2, \\
    V_{s\phi} &= V_{\phi s} = -\frac{M\pi^2}{2}, \quad
    V_{\phi\tau} = V_{\tau\phi} = - GLbt_f \pi^2\lambda^2, \quad
    V_{s\tau} = V_{\tau s} = V_{t\tau} = V_{\tau t} = V_{t\phi} =
    V_{\phi t} = 0,
  \end{split}
\end{equation}
which when substituted into $\mathbf{V}_{ij}$ with the singular
condition at the critical point $\mathrm{C}$, where $M=\Mcr$, gives:
\begin{equation}
  \Mcr = \frac{\pi}{L} \sqrt{\frac{EI_y GJ}{1+s}} \sqrt{ 1 +
      \frac{1}{(1+s)}\frac{\pi^2}{L^2} \frac{EI_w}{GJ}}.
\label{eq:Mcr_new}
\end{equation}
This new expression for $\Mcr$ replaces equation (\ref{eq:Mcr}) and is
used subsequently. The term, $s=E\pi^2 \psi^2/12G$, accounts for the
non-zero shear distortion of both flanges, which tends to zero if $G$
or $L$ become large; this is an entirely logical result reflecting the
difference between Timoshenko and Euler--Bernoulli beam theories
\cite{Shear}. However with $s>0$, the above expression gives values
that are marginally below those given by the classical critical moment
given in equation (\ref{eq:Mcr}).

\subsubsection{Equilibrium equations}
\label{sec:equil}

The total potential energy $V$ with the rescaled variables can be
written thus:
\begin{equation}
  \label{eq:lagrangian}
  V = \int_0^2 \mathcal{L}(\tw'',\tw',\tw,\tu',\tu;\tz) \D \tz.
\end{equation}
Equilibrium equations are found where $V$ is stationary and this is
established from setting the first variation of $V$ (or $\delta V$) to
zero, where:
\begin{equation}
  \label{eq:1stvar}
  \begin{split}    
    \delta V &= \int_0^2 \left[ \frac{\partial\mathcal{L}}{\partial
        \tw''} \delta \tw'' + \frac{\partial\mathcal{L}}{\partial \tw'}
      \delta \tw' + \frac{\partial\mathcal{L}}{\partial \tw} \delta \tw +
      \frac{\partial\mathcal{L}}{\partial \tu'} \delta \tu'
      + \frac{\partial\mathcal{L}}{\partial \tu} \delta \tu \right] \D \tz \\
    &= \left\{ \frac{\partial\mathcal{L}}{\partial \tw''} \delta \tw'
      +\left[ \frac{\partial\mathcal{L}}{\partial \tw' } -
        \frac{\Dx}{\Dx \tz} \left(\frac{\partial\mathcal{L}}{\partial
            \tw''}\right) \right] \delta \tw +
      \frac{\partial\mathcal{L}}{\partial \tu'}\delta \tu
    \right\}_0^2\\
    & \qquad + \int_0^2 \left\{ \left[ \frac{\Dx^2}{\Dx \tz^2}
        \left(\frac{\partial\mathcal{L}}{\partial \tw''}\right) -
        \frac{\Dx}{\Dx \tz} \left(\frac{\partial\mathcal{L}}{\partial
            \tw'} \right) + \frac{\partial\mathcal{L}}{\partial \tw}
      \right] \delta \tw + \left[ \frac{\partial\mathcal{L}}{\partial \tu}
        - \frac{\Dx}{\Dx \tz} \left(\frac{\partial\mathcal{L}}{\partial
            \tu'} \right) \right] \delta \tu \right\} \D \tz.
\end{split}
\end{equation}
The term in the integral has to vanish for all $\delta w$ and $\delta
u$, which gives two coupled nonlinear ordinary differential equations:
\begin{equation}
  \label{eq:ode_w}
  \begin{split}
    & \tw'''' - \frac{6(1-\nu)}{\psi^2} \tw''
    -\frac{3GL^2t_f}{8 D \psi^2} \tw \left[ \tu' + \frac23 (\tw'^2 + \tw \tw'')
      + \left[q_s - q_t + \lambda (q_\phi - q_\tau ) \right]
      \frac{\pi^2 \psi}{2} \sin \frac{\pi \tz}{2} \right]\\
    & \quad
    + \frac{3EL^2t_f}{8D}
    \biggl[ (q_t + q_\tau \lambda) \frac{\pi^2 \psi}{4} \left( \sin
      \frac{\pi \tz}{2} \tw''
      + \frac{\pi}{2} \cos \frac{\pi \tz}{2} \tw' \right) - \frac12
    (\tu'' \tw'+ \tu' \tw'') - \frac35 \tw'^2 \tw''\biggr] = 0, 
  \end{split} 
\end{equation}
\begin{equation}
  \label{eq:ode_u}
  \tu'' + \frac34 \tw'\tw'' + \frac{\pi^2}{4s} \left\{ \pi \psi \cos \frac{\pi
      z}{2} \left[ \left( q_s - q_t + \lambda ( q_\phi - q_\tau) \right) -
    s(q_t + q_\tau \lambda ) \right] - \left(\tu+ \frac12 \tw \tw'
    \right) \right\} = 0,
\end{equation}
subject to the following boundary conditions which arise from
minimizing the terms outside the integral in equation
(\ref{eq:1stvar}):
\begin{align}
  \label{eq:pinned}
  \tw(0) = \tw''(0) = \tw(2) = \tw''(2) &= 0,\\
  \label{eq:endstrain}
  \tu'(0) + \frac38 \tw'^2 (0) + \frac{3M}{Ebht_f} &= 0,\\
  \label{eq:symmsec}
  \tu(1) = \tw'(1) = \tw'''(1) &= 0,
\end{align}
where equation (\ref{eq:pinned}) refers to pinned boundaries, equation
(\ref{eq:endstrain}) refers to the end strain condition, and equation
(\ref{eq:symmsec}) refers to reflective symmetry of $w$ and
antisymmetry of $u$ about the midspan respectively. The symmetry
conditions are particularly pertinent when LTB occurs simultaneously
with or before local buckling owing to the sinusoidal distribution of
$\phi$ forcing the maximum bending stress to be located at midspan.

Other equilibrium equations can be obtained by minimizing the energy
with respect to the generalized coordinates $q_s$, $q_\phi$, $q_t$ and
$q_\tau$:
\begin{align}
  \label{eq:Vqs}
  \frac{\partial V}{\partial q_s} &= \frac{G\pi^2 Lbt_f}{4}
  \int_0^2 \biggl\{ \left[  4 (q_s - q_t)
    \cos \frac{\pi \tz}{2} - \frac{1}{\pi\psi} \left( \tu + \frac12
      \tw \tw' \right) \right] \cos \frac{\pi \tz}{2}
  - \frac{2M}{GLbt_f} q_\phi \sin^2 \frac{\pi \tz}{2} \biggr\} \D \tz= 0,\\
  \label{eq:Vqphi}
  \frac{\partial V}{\partial q_\phi} &= \frac{G\pi^2 bht_f}{2}
  \int_0^2 \biggl\{ \left[ \left( \lambda ( q_\phi - q_\tau ) +
    \frac{J}{Lbht_f} q_\phi \right) \cos \frac{\pi \tz}{2}-
  \frac{1}{4\pi\psi} \left( \tu + \frac12 \tw \tw' \right) \right] \cos
  \frac{\pi \tz}{2} \\
  \nonumber & \qquad \qquad \qquad \qquad
  - \frac{M}{Gbht_f} q_s \sin^2 \frac{\pi \tz}{2} \biggr\} \D \tz = 0,\\
  \label{eq:Vqt}
  \frac{\partial V}{\partial q_t} &= \frac{E\pi^2 Lbt_f}{4}
  \int_0^2 \biggl\{ \left[ 2 q_t \pi^2 \psi
      \sin \frac{\pi \tz}{2} - \left( \tu' + \frac38 \tw'^2 \right)
    \right] \frac{\psi}{6} \sin \frac{\pi \tz}{2} \\ \nonumber
    & \qquad \qquad \qquad \qquad
    + \left[ \frac{1}{\pi\psi}
      \left(\tu + \frac12 \tw \tw' \right) - 4 (q_s - q_t) \cos
      \frac{\pi \tz}{2} \right] \frac{G}{E} \cos \frac{\pi \tz}{2}
    \biggr\} \D \tz = 0,\\
    \label{eq:Vqtau}
  \frac{\partial V}{\partial q_\tau} &= \frac{E\pi^2 bh t_f}{8} \int_0^2
  \biggl\{ \left[ 2q_\tau \lambda \pi^2 \psi 
    \sin \frac{\pi \tz}{2} - \left( \tu' + \frac38 \tw'^2 \right)
  \right] \frac{\psi}{6} \sin \frac{\pi \tz}{2} \\ \nonumber
  & \qquad \qquad \qquad \qquad
  + \left[ \frac{1}{\pi\psi}
    \left(\tu + \frac12 \tw \tw' \right) - 4 \lambda (q_\phi - q_\tau) \cos
    \frac{\pi \tz}{2} \right] \frac{G}{E} \cos \frac{\pi \tz}{2} \biggr\}
  \Dx \tz = 0.
\end{align}

\section{Physical experiments}
\label{sec:expts}

\subsection{Specimens and procedure}

\begin{figure}[htbp]
  \centerline{\psfig{figure=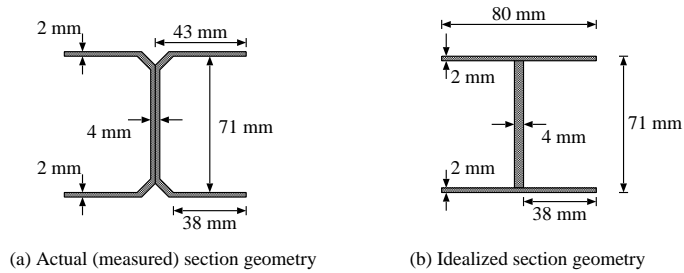,width=90mm}}
  \caption{Cross-section of steel beams tested in the experimental
    programme. (a) Actual section constructed from two channel
    sections; (b) idealized section geometry derived from
    experiments.}
  \label{fig:exptsection}
\end{figure}

A series of physical experiments were conducted on I-beams fabricated
by spot-welding thin-walled channel sections, made from cold-formed
steel, back-to-back. The key material properties were measured to be
thus: Young's modulus $E=205~\mathrm{kN/mm^2}$, Poisson's ratio
$\nu=0.3$ and the yield stress
$\sigma_\mathrm{y}=290~\mathrm{N/mm^2}$. The channel sections were $75
\times 43 \times 2~\mm$ in terms of depth, width and thickness
respectively -- see Figure \ref{fig:exptsection}(a).
The actual geometry (including corner radii \emph{etc}.)\ was
converted into an idealized I-section comprising only flat plate
elements based on the mean slope of the linear regions of the measured
load versus maximum bending displacement curves from the tests. The
dimensions of $h$ and $b$ were hence adjusted slightly such that a
meaningful comparison with the theory could be made -- see Figure
\ref{fig:exptsection}(b).  Each beam, the idealized properties of
which are given in Table \ref{tab:expts},
\begin{table}[htb]
  \centering
  \renewcommand\arraystretch{1.0}
  \begin{tabular}{c|c||c|c}
    Geometric Property & Value & Cross-Section Property & Value\\
    \cline{1-4}
    $b$ & $80~\mm$ & $I_x$ & $4.81 \times 10^5~\mm^4$\\
    $h$ & $71~\mm$ & $I_y$ & $1.71 \times 10^5~\mm^4$\\
    $t_f$ & $2~\mm$ & $I_w$ & $2.16 \times 10^8~\mm^6$\\
    $t_w$ & $4~\mm$ & $J$ & $1.86 \times 10^3~\mm^4$\\
  \end{tabular}
  \caption{Idealized section properties of the experimental samples.}
  \label{tab:expts}
\end{table}
had an overall length of 4 metres, was tested under four-point bending
with a specified buckling length $L_e$ given in Table
\ref{tab:Le}
\begin{table}[htb]
  \centering
  \begin{tabular}{c|c|c|c|c}
    Test & Effective length $L_e~(\mm)$ &
    $\sigma^\mathrm{C}_\mathrm{Local}/\sigma^\mathrm{C}_\mathrm{LTB}$ &
    $\sigma^\mathrm{C}~(\mathrm{N/mm^2})$
    & Critical mode\\
    \cline{1-5}
    $1$ & $3200$ & 1.19 & 186 & LTB\\
    $2$ & $3200$ & 1.19 & 186 & LTB\\
    $3$ & $3000$ & 1.10 & 202 & LTB\\
    $4$ & $2750$ & 0.98 & 221 & Local (marginally)\\
    $5$ & $2500$ & 0.87 & 221 & Local\\
    $6$ & $2250$ & 0.75 & 221 & Local\\
  \end{tabular}
  \caption{Buckling lengths for each beam test which reflect the
    symmetric position of the lateral restraints; these lengths were
    chosen such that
    $\sigma^\mathrm{C}_\mathrm{Local}/\sigma^\mathrm{C}_\mathrm{LTB} =
    [0.75,1.20]$ with one beam (Test 4) triggering both modes approximately
    simultaneously. }
  \label{tab:Le}
\end{table}
that was controlled by an adjustable pair of lateral restraints (see
Figure \ref{fig:exptsetup} in \S\ref{sec:rig}). The critical mode was
determined by comparing the strong axis bending stress when $M=\Mcr$,
thus:
\begin{equation}
  \label{eq:LTB_sigma}
  \sigma^\mathrm{C}_\mathrm{LTB} =  \frac{\Mcr y_\mathrm{max}}{I_x} =
  \frac{\Mcr h}{2 I_x},
\end{equation}
with the critical stress of plate buckling being given by equation
(\ref{eq:local_sigma}).

\subsection{Testing rig}
\label{sec:rig}

\begin{figure}[htb]
  \centerline{\psfig{figure=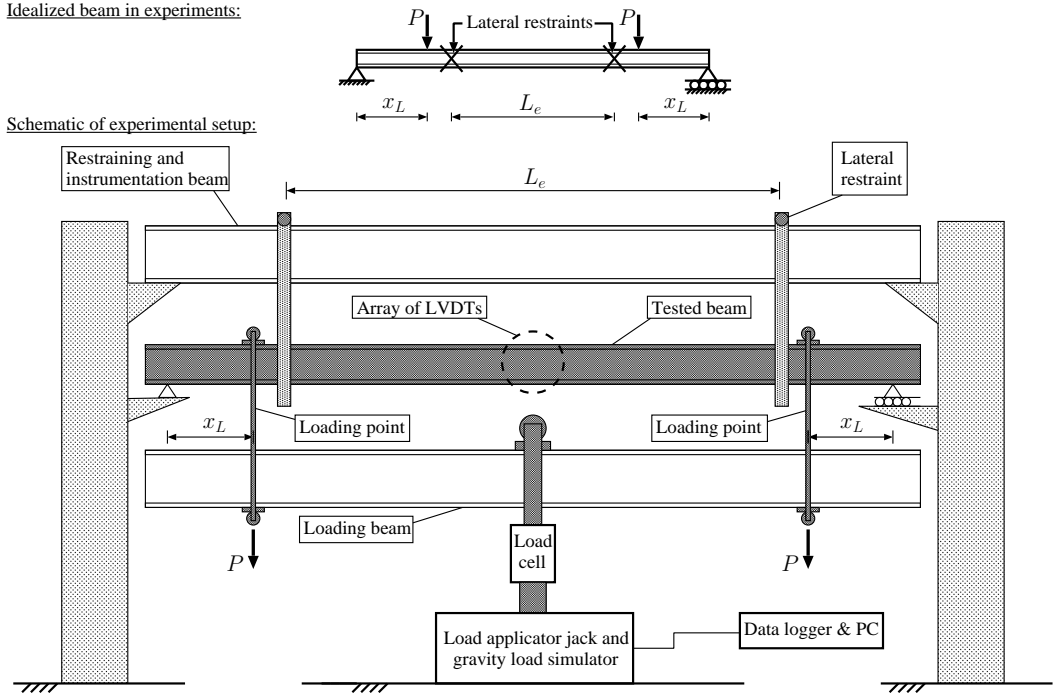,width=140mm}}
  \caption{Experimental rig designed to test fabricated I-beams under
    four-point bending. Although each beam was $4~\metre$ in length,
    the distance between the pinned supports was in fact
    $3.8~\metre$. In each test, the lateral restraints were adjusted
    symmetrically to obtain the desired buckling length $L_e$.}
  \label{fig:exptsetup}
\end{figure}

\begin{figure}[htbp]
  \centerline{\psfig{figure=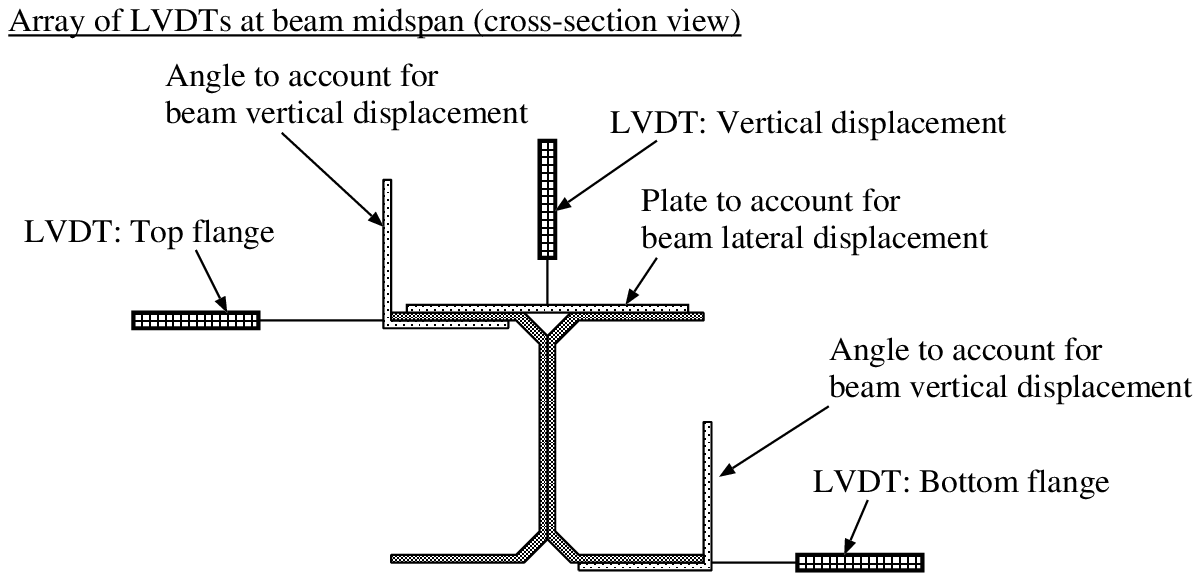,width=100mm}}
  \caption{Cross-section of the tested beam showing the relative
    positions of the LVDTs. Angles and plates were clamped as shown to
    both flanges such that the overall beam displacement still allowed
    the relative flange and vertical displacements to be evaluated.}
  \label{fig:LVDT}
\end{figure}

A schematic of the experimental setup along with an idealized
representation is shown in Figure \ref{fig:exptsetup}.
The total applied load $2P$ was split into two point loads each of $P$
applied at a distance $x_L$ from the end supports. Hence, from simple
statics, the uniform moment $M$ between the two loading points was
$Px_L$.  The loading was displacement controlled; it was applied with
a hand-operated hydraulic jack in conjunction with a gravity load
simulator, a mechanism that adjusted the position of the load
application relative to the deflecting beam such that the applied load
remained vertical. Since the jack was hand operated, the displacement
was applied in short controlled increments but it did mean that
dynamic loads were inevitable to a small extent.  At midspan, the
vertical displacement of the beam and lateral displacements of the
flanges were measured using linear variable differential transformers
(LVDTs); the locations of which are presented in Figure
\ref{fig:LVDT}.

The large displacements and dynamic behaviour of unstable
post-buckling, even with displacement control, meant that sometimes
the LVDTs measuring the displacement of the greater displacing top
flange -- see Figure \ref{fig:Isection}(b) -- went out of range very
quickly after the secondary instability was triggered. However, there
were no such problems associated with the bottom flange measurements
and so these were the primary values used for comparison purposes,
since the theoretical model gives both $u_s$ and $\phi$ directly.

\section{Numerical results and validation}

\subsection{Cellular buckling}

\begin{figure}[htb]
  \centerline{\psfig{figure=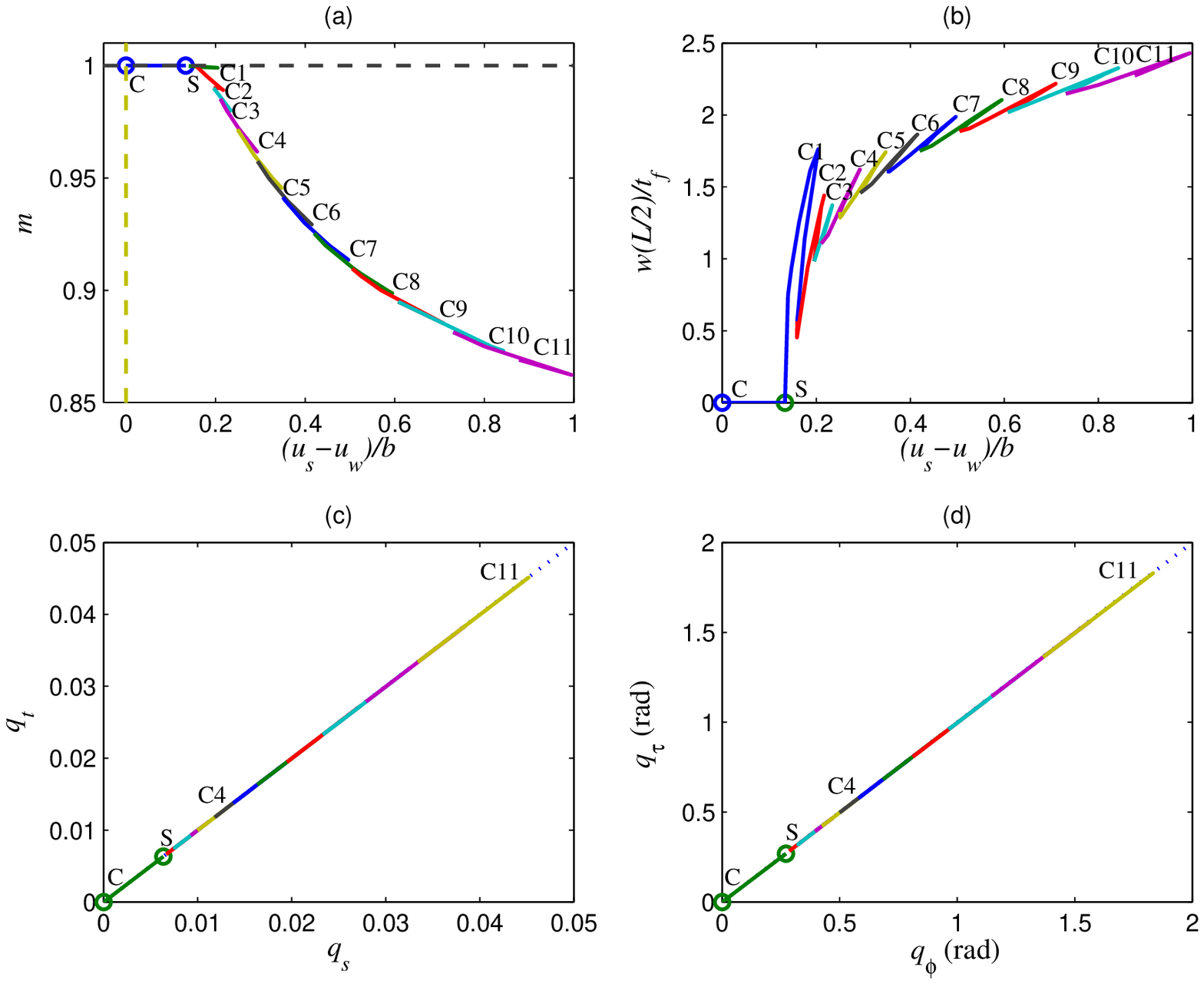,width=140mm}}
  \caption{Numerical equilibrium paths for test specimen 1. Graphs of
    (a) normalized moment ratio $m$ and (b) local buckling
    displacement amplitude $w_\mathrm{max}/t_f$ versus lateral
    displacement of the bottom flange are shown.  Points $\mathrm{C}$
    and $\mathrm{S}$ show the critical and secondary bifurcation
    points respectively. Note that the sequence of paths, separated by
    snap-backs that mark the appearance of a new local buckling
    ``cell'', are denoted as ``$\mathrm{Cn}$'' where $\mathrm{n}$ is
    the cell number. The graphs in (c) and (d) show the relationships
    between the generalized coordinates defining LTB during
    interactive buckling; note that the dotted lines represent the
    relationship between the respective quantities assuming
    Euler--Bernoulli bending, which show that the developed shear
    strains are small.}
  \label{fig:Mlat_tf2_tw4_b80_h75_L3200}
\end{figure}

\begin{figure}[htb]
  \centerline{\psfig{figure=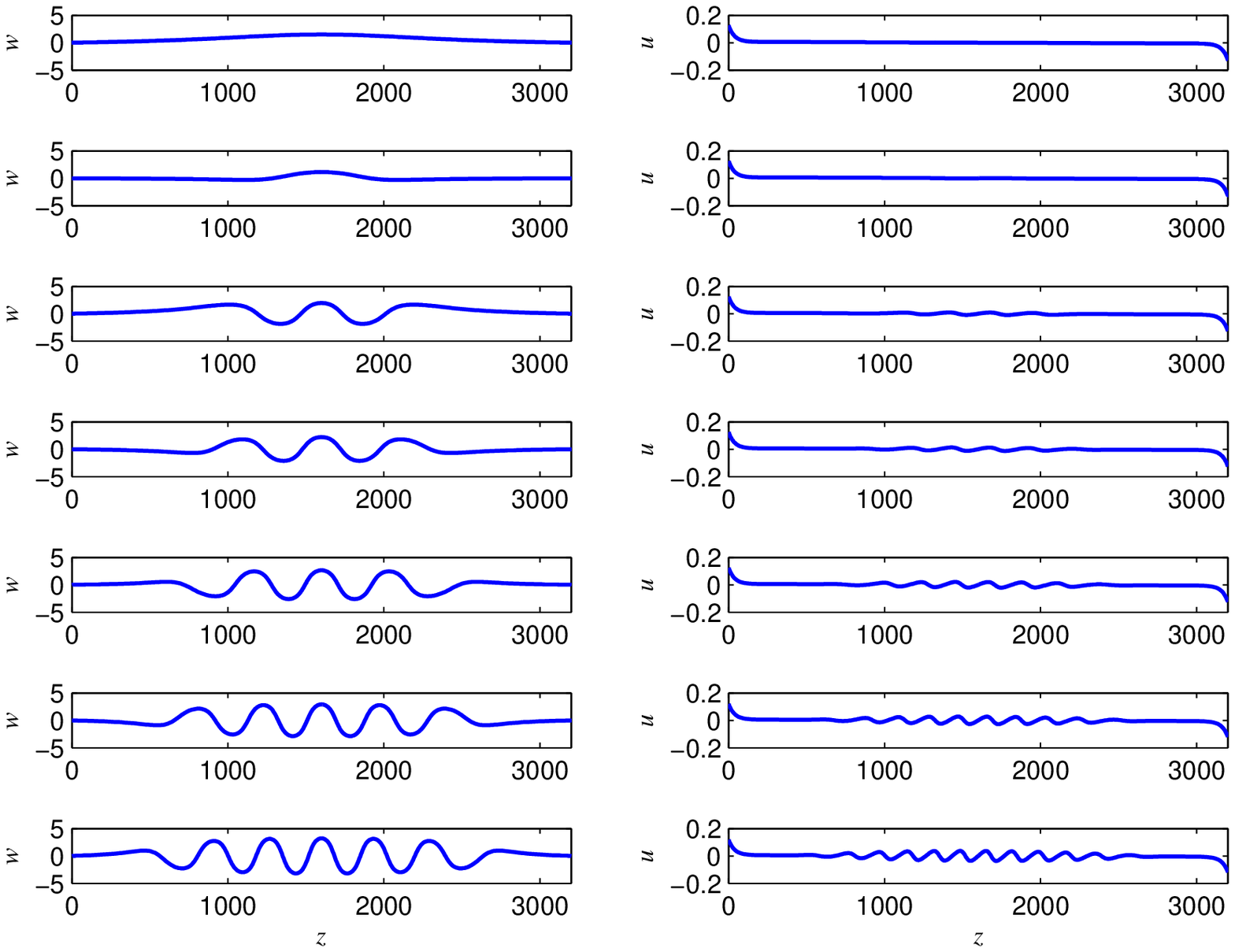,width=140mm}}
  \caption{Numerical solutions of the displacements of local buckling
    for the tip ($x=-b/2$) of the vulnerable flange of test specimen
    1: out-of plane displacement $w$ (left) and in-plane displacement
    $u$ (right).  Individual solutions on equilibrium paths C1 to C7
    (defined in Figure \protect\ref{fig:Mlat_tf2_tw4_b80_h75_L3200})
    are shown in sequence from top to bottom respectively.  All
    dimensions are in millimetres.}
  \label{fig:wu_vs_z}
\end{figure}

\begin{figure}[htbp]
  \centering
  \subfigure[C1: $m=0.9998$]{%
    \psfig{figure=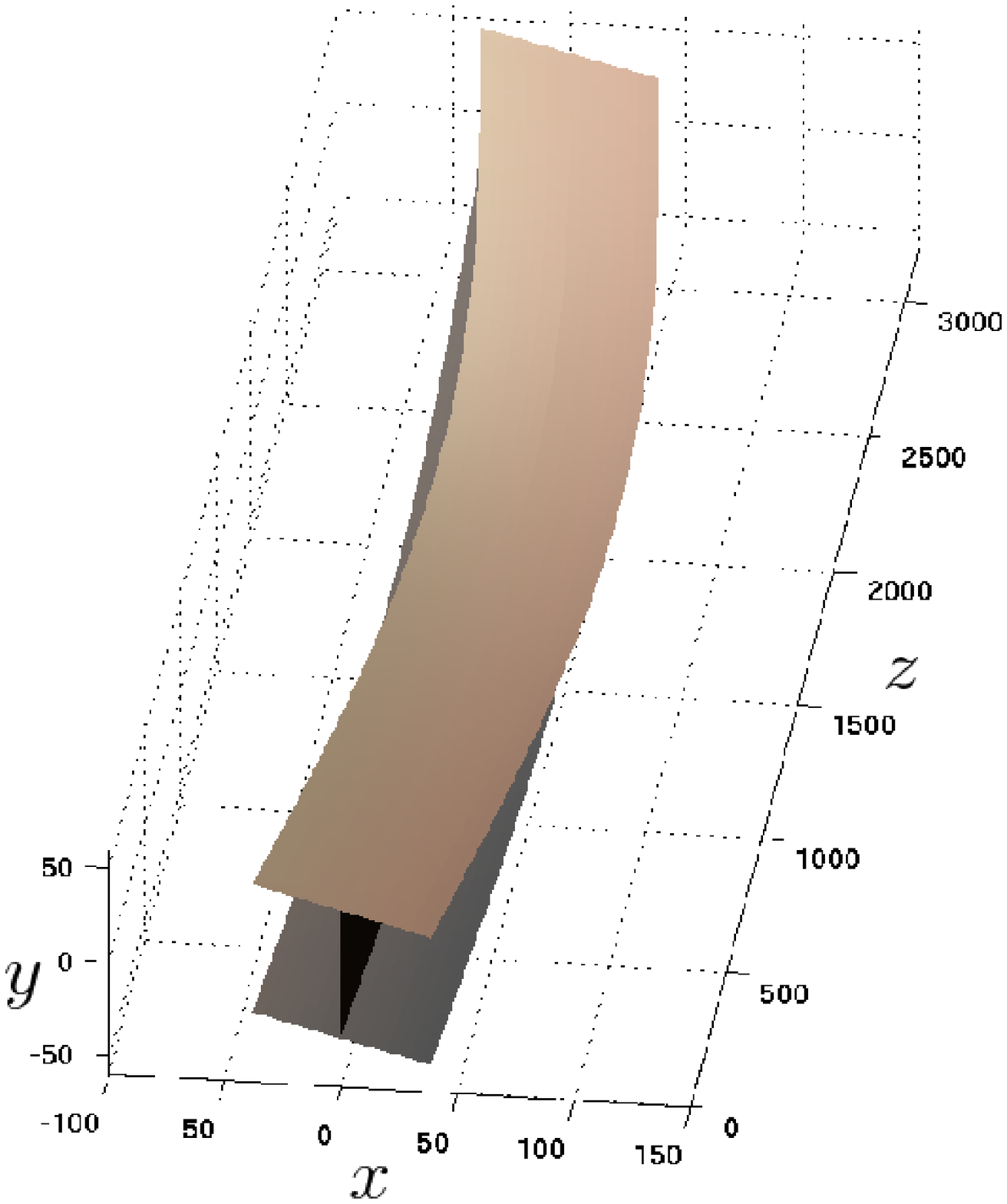,height=80mm}}\quad
  \subfigure[C3: $m=0.9900$]{%
    \psfig{figure=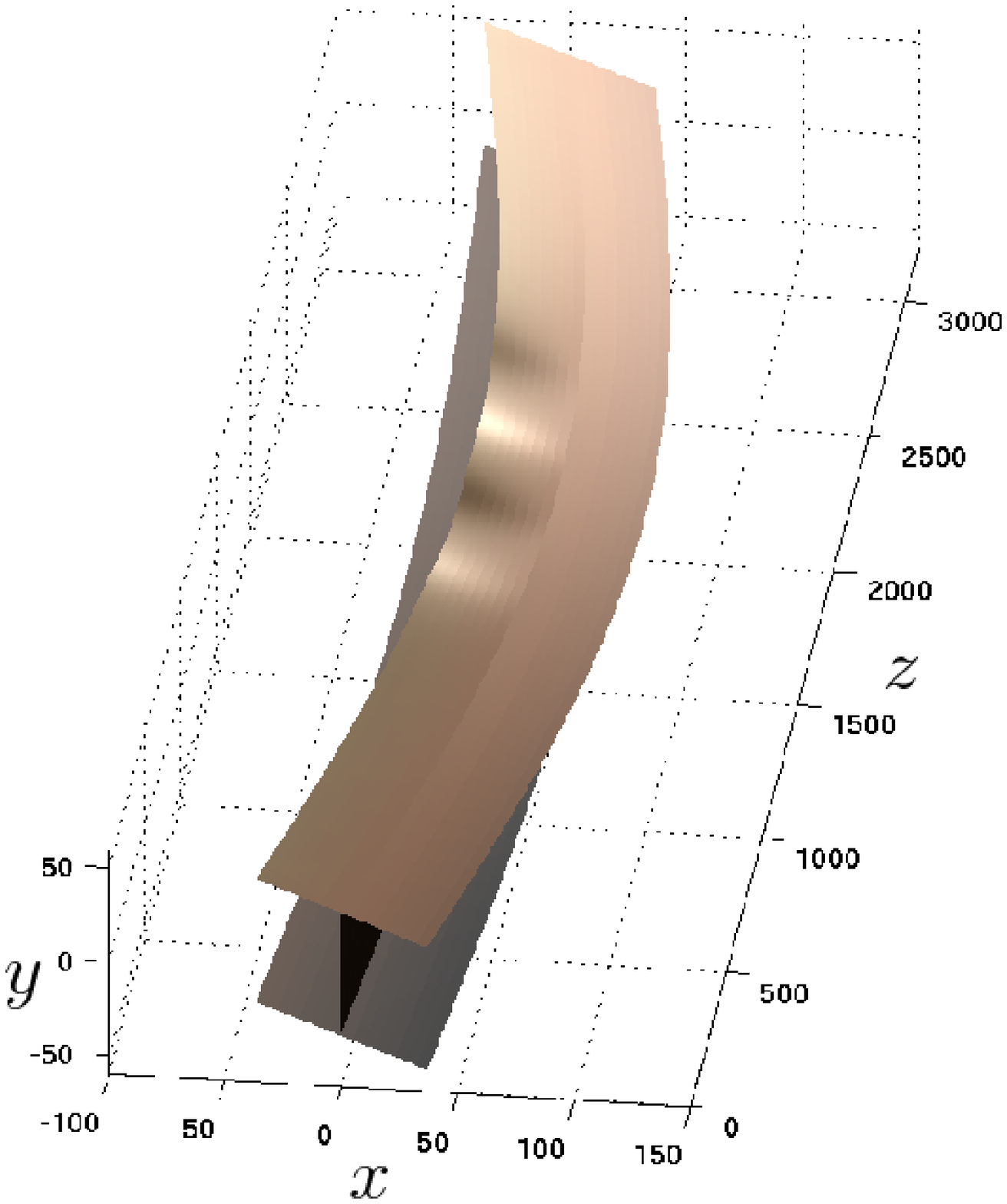,height=80mm}}\\
  \subfigure[C5: $m=0.9718$]{%
    \psfig{figure=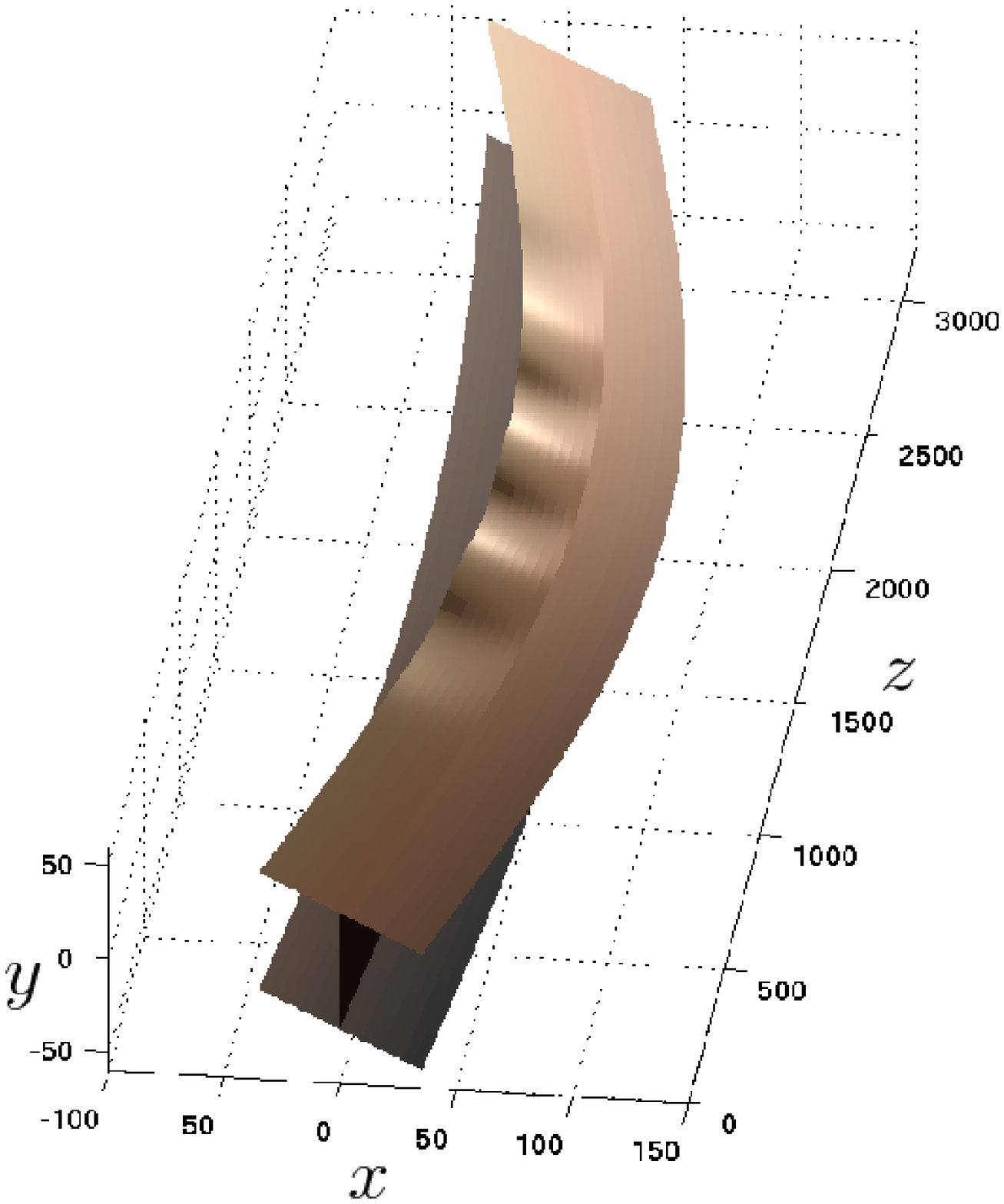,height=80mm}}\quad
  \subfigure[C7: $m=0.9414$]{%
    \psfig{figure=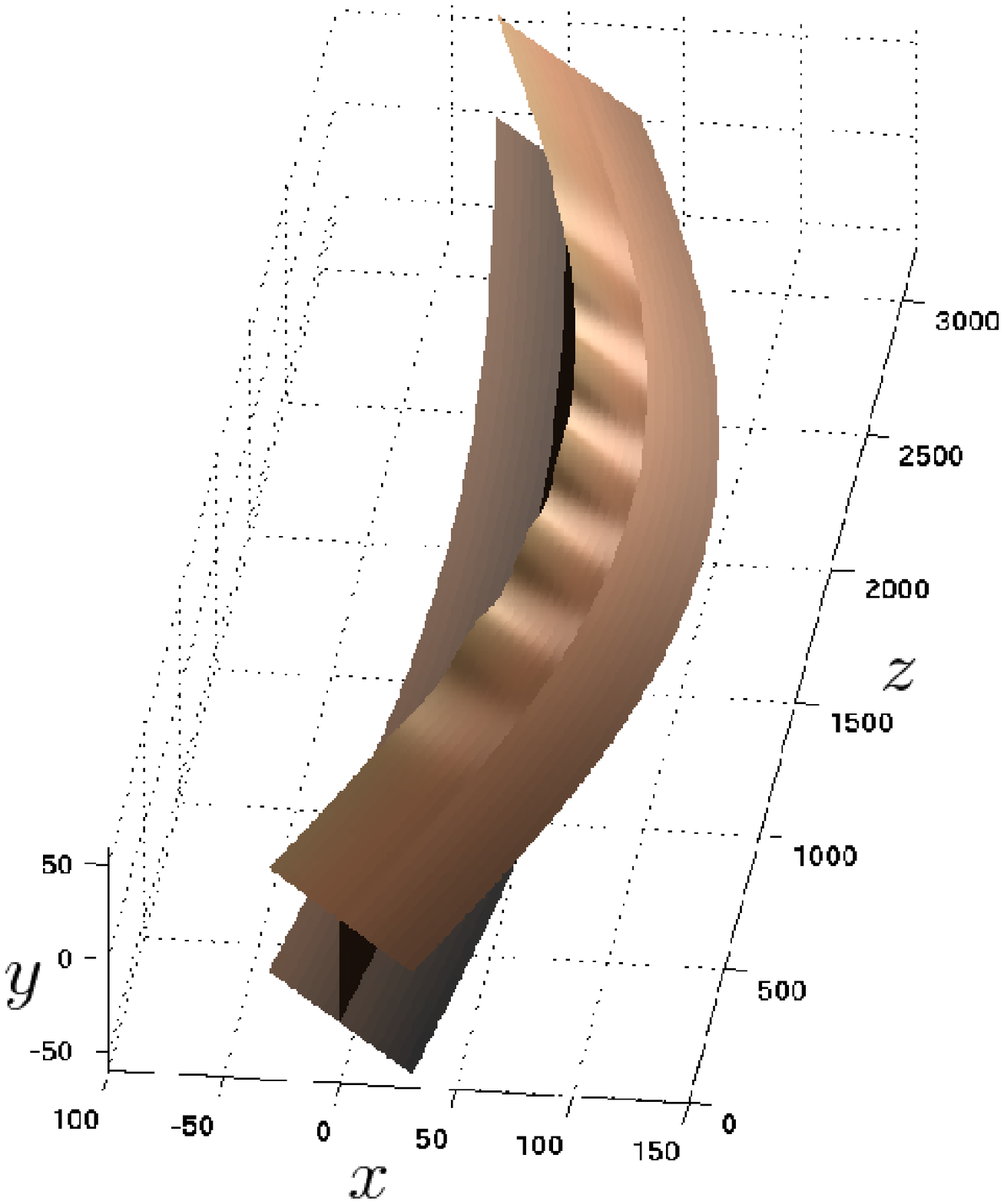,height=80mm}}
  \caption{Numerical solutions of the system of equilibrium equations
    visualized on a representation of the actual beam used in test
    1. All deformation components: $w$, $u$, $u_s$, $\phi$, $u_{tb}$
    and $u_{tt}$ are included with all dimensions in millimetres. The
    results are shown for individual points on paths C1, C3, C5 and C7
    (Figure \protect\ref{fig:Mlat_tf2_tw4_b80_h75_L3200}) with the
    specific moment ratio $m$ given. Note how the local buckling mode
    develops and how the ``wavelength'' of the local buckling profile
    within the central portion of the flange in more compression
    changes as more cells develop.}
  \label{fig:3d_beams}
\end{figure}

The system of nonlinear ordinary differential equations
(\ref{eq:ode_w})--(\ref{eq:ode_u}), subject to boundary conditions
from equations (\ref{eq:pinned})--(\ref{eq:symmsec}) and integral
equations from (\ref{eq:Vqs})--(\ref{eq:Vqtau}) are solved using the
well-known and tested numerical continuation package \textsc{Auto}
\cite{Auto2007}. For illustrative purposes, Figures
\ref{fig:Mlat_tf2_tw4_b80_h75_L3200}--\ref{fig:3d_beams}
present results from the variational model for test specimen 1. Figure
\ref{fig:Mlat_tf2_tw4_b80_h75_L3200} shows a plot of the (a)
normalized moment ratio $m$, which is defined as the ratio $M / \Mcr$,
and (b) the normalized local buckling displacement amplitude
($w(L/2)/t_f$) of the vulnerable part of the compression flange versus
the normalized lateral displacement of the bottom flange,
$(u_s-u_w)/b$. The graphs in (c) and (d) show the relationships
between the ``sway'' and ``tilt'' components of the weak axis
centroidal displacement ($q_s$ and $q_t$) and the torsional angle
($q_\phi$ and $q_\tau$). A dotted line is superimposed on these graphs
to show the Euler--Bernoulli assumption, where the sway and tilt
amplitudes would be equal; this shows that the shear strains developed
are small but not zero. Moreover, Figures
\ref{fig:Mlat_tf2_tw4_b80_h75_L3200}(a) and (b) show a series of paths
separated by a sequence of snap-backs with Figure \ref{fig:wu_vs_z}
presenting detailed graphs showing the corresponding numerical
solutions beyond individual snap-backs for the local buckling
functions.  A distinctive pattern is clearly seen to emerge where the
response passes from one path to the next, \emph{i.e.} from C1 to C2
to C3 and so on, in which each new path reveals a new local buckling
displacement peak or trough.  A selection of 3-dimensional
representations of the beams using the solutions for the paths C1, C3,
C5 and C7 are presented in Figure \ref{fig:3d_beams},
which include all components of LTB ($u_s$, $\phi$, $u_{tb}$ and
$u_{tt}$) and local buckling ($w$ and $u$).

As the response advances to path C11, which has torsional rotations
that are well beyond the scope of the model in terms of geometric
considerations, the local buckle pattern is all but periodic and
further loading would restabilize the system globally, assuming no
permanent deformation has taken place. This global restabilization
would occur as a result of the boundaries confining the spread of the
buckling profile any further. Of course, if plasticity were present in
the flange then any restabilization would be less significant and
displacements would lock into plastic hinges.

The phenomenon demonstrated currently and described above, where a
sequence of snap-backs causes a progressive change from an initially
localized post-buckling mode to periodic, has been termed in the
literature as \emph{cellular buckling} \cite{nondyn} or \emph{snaking}
\cite{BurkeKnobloch2007}. It has been found to be prevalent in systems
where there is progressive destabilization and subsequent
restabilization \cite{wadee_bassom_jmps_2000,Peletier_SIAM_2001}, such
as in cylindrical shell buckling \cite{HLC98,HuntLordPeletier03} and
kink banding in confined layers \cite{MAW_jsg,MAW_jmps05}. In the
fundamental studies concerning the model strut on a nonlinear
foundation, the load oscillates about the Maxwell load, where the
buckling modes progressively transform from localized (homoclinic)
profiles to a periodic mode in a heteroclinic connection
\cite{cjb_gwh_rk_prsa_2001,wadee_bassom_coman_physD_2002}. The
oscillation in the strut model is attributed to the combination of
nonlinearities in the foundation that have softening and hardening
properties. In the present context, as in the case for sandwich struts
\cite{nondyn}, the destabilization is derived from the interaction of
instability modes with the restabilization arising from the inherent
stretching that occurs during plate buckling due to large deflections,
which accounts for its significant post-buckling stiffness
\cite{Koiter76}. Moreover, since the moment ratio $m$ has a decreasing
trend rather than oscillating about a fixed value, it is suggested
that the destabilization is inherently more severe than the
restabilization for the present case.

\subsection{Comparison with existing experiments}
\label{sec:existing}

\begin{figure}
  \centerline{\psfig{figure=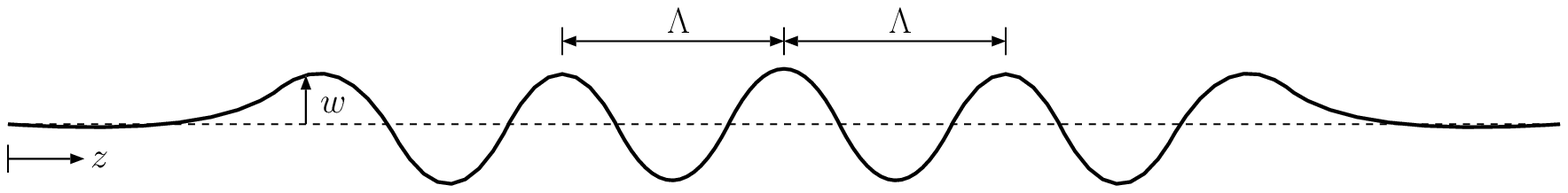,width=120mm}}
  \caption{Definition of local buckling wavelength $\Lambda$ from
    results for $w$ from the variational model.}
  \label{fig:waves}
\end{figure}

Work conducted by Cherry \cite{cherry_ltb_local_1960}, which focused
on the overall buckling strength of beams under bending that had
locally buckled flanges, presented a series of test results and
proposed a theoretical estimate of the post-buckling strength. The
theoretical approach was based on the effective width of the locally
buckled flange which originated in
\cite{vk_sechler_donnell_1932}. However, the model presented in
\cite{cherry_ltb_local_1960} was limited because of the assumption
that both outstands of the compression flange behaved
symmetrically. Nevertheless, the tests that were presented therein
provide valuable data for the wavelengths of the local buckling mode
in the compression flanges that were measured for four separate
doubly-symmetric I-beams, with properties as presented in Table
\ref{tab:cherrybeams}; the data are used for comparison purposes in
the current study.
\begin{table}[hbtp]
  \centering
  \begin{tabular}{c|c|c|c|c|c}
    Beam & $h~(\mm)$ & $b~(\mm)$ & $t_w~(\mm)$ & $t_f~(\mm)$ &
    $E~(\mathrm{kN/mm^2})$ \\
    \cline{1-6}
    A & 72.3 & 76.6 & 3.68 & 1.60 & 66.47 \\
    B & 74.0 & 76.6 & 3.78 & 2.11 & 64.53 \\
    C & 72.2 & 76.3 & 5.13 & 1.89 & 64.43 \\
    D & 71.5 & 75.9 & 4.65 & 1.90 & 66.19 \\
  \end{tabular}
  \caption{Cross-section properties of beams tested and results presented in
    \protect\cite{cherry_ltb_local_1960}. Three different lengths of beams
    were tested under bending for each cross-section (A--D) ranging
    from $1~\metre$ to $2~\metre$.}
  \label{tab:cherrybeams}
\end{table}
Since the buckling mode predicted by the current analytical model is
not necessarily periodic, but tends to approach this quality in the
far post-buckling range, comparisons between the tests in
\cite{cherry_ltb_local_1960} and the current model can be made when
the profile for $w$ exhibits periodicity throughout the beam
length. Figure \ref{fig:waves}
shows how the wavelength is defined from the post-buckling mode that
has a central portion which is close to periodic. Table
\ref{tab:cherry_compare}
\begin{table}[hbtp]
  \centering
  \begin{tabular}{c||c||c|c||c}
    Beam & Test wavelength \protect\cite{cherry_ltb_local_1960} &
    \multicolumn{2}{|c||}{Model wavelength $\Lambda$ ($\mm$)} &
    Error range\\
    & $(\mm)$ & Minimum & Maximum & (\%) \\
    \cline{1-5}
    A & 126.2 & 127.0 & 133.4 & $+0.6 \rightarrow +5.7$ \\
    B & 124.0 & 153.0 & 156.7 & $+23.4 \rightarrow +26.4$ \\
    C & 108.5 & 103.1 & 132.6 & $-4.9 \rightarrow +22.3$ \\
    D & 109.7 & 102.1 & 141.2 & $-7.0 \rightarrow +28.7$ \\
  \end{tabular}
  \caption{Comparisons of buckling wavelengths between the tests in
    \protect\cite{cherry_ltb_local_1960} and the current model.} 
  \label{tab:cherry_compare}
\end{table}
shows the range of wavelengths obtained from the numerical solution of
the system of equilibrium equations presented in \S\ref{sec:equil}.
The current model was run for a range of lengths between
$1$--$2~\metre$ since this was the range for which the vast majority
of tests presented in \cite{cherry_ltb_local_1960} were
conducted. Apart from the beams with cross-section B, the comparisons
are very encouraging; the discrepancies between the model and the
tests are attributed to boundary effects affecting the results of the
analytical model. It has been seen in the cellular buckling results
earlier in this section (Figures
\ref{fig:Mlat_tf2_tw4_b80_h75_L3200}--\ref{fig:3d_beams}) that, as
each buckling cell develops, the buckling ``wavelength'' $\Lambda$,
see Figure \ref{fig:waves}, drops until the buckling profile
eventually tends to true periodicity and the moment $M$ tends to a
constant. For the numerical results from the current model that
overestimated the wavelength, lack of convergence became an issue and
the local buckling profile $w$ was still showing remnants of the
decaying tails near the boundaries, which are the signatures of
homoclinic behaviour. Hence, those particular comparisons are perhaps
not entirely representative of the actual response predicted by the
model.

\subsection{Results from current experiments}
\label{sec:exptres}

\begin{figure}[htb]
\centering
\subfigure[Test 1: Close up of local buckling waves]{%
  \psfig{figure=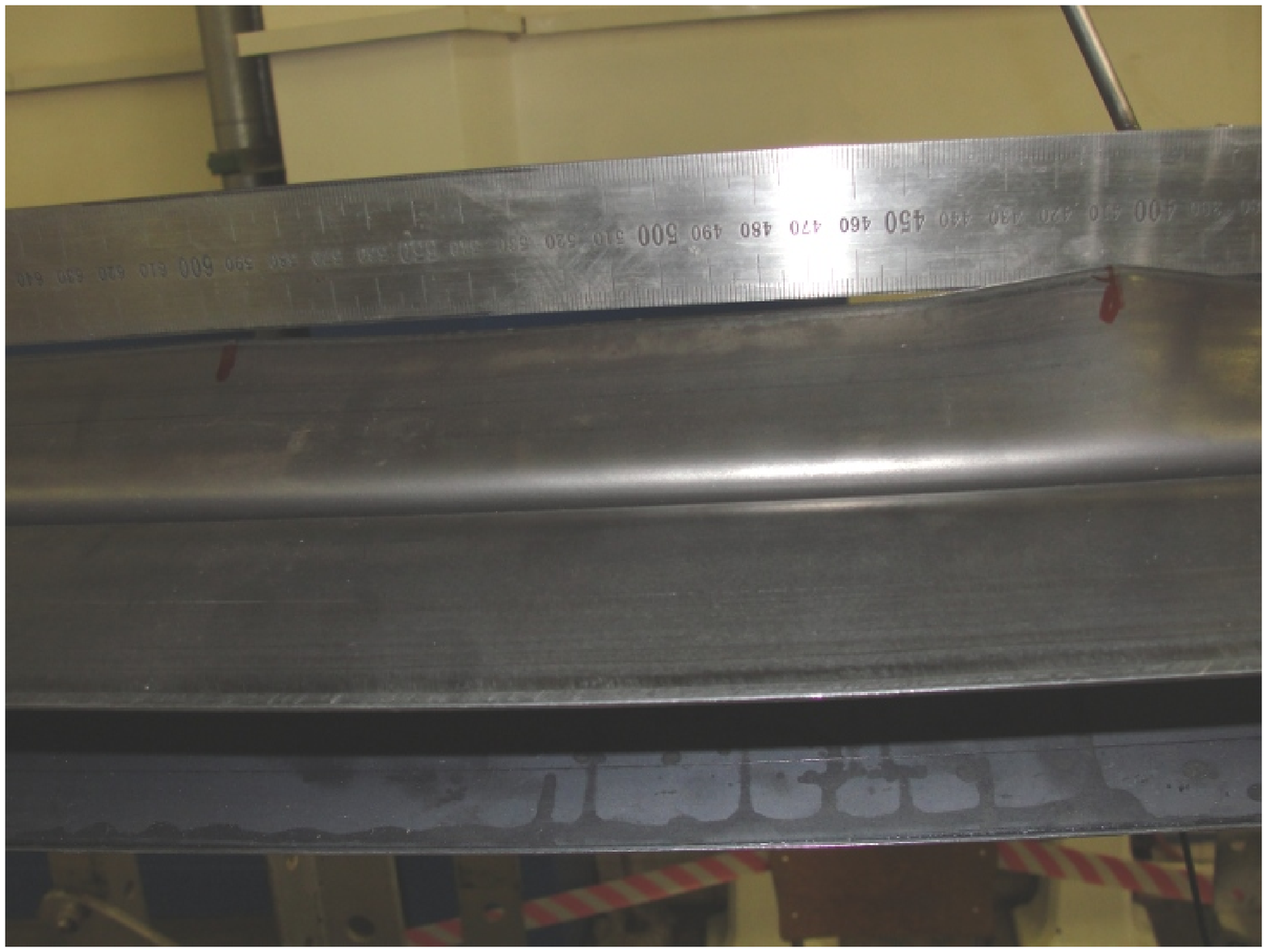,width=74mm}}
\subfigure[Test 3]{%
  \psfig{figure=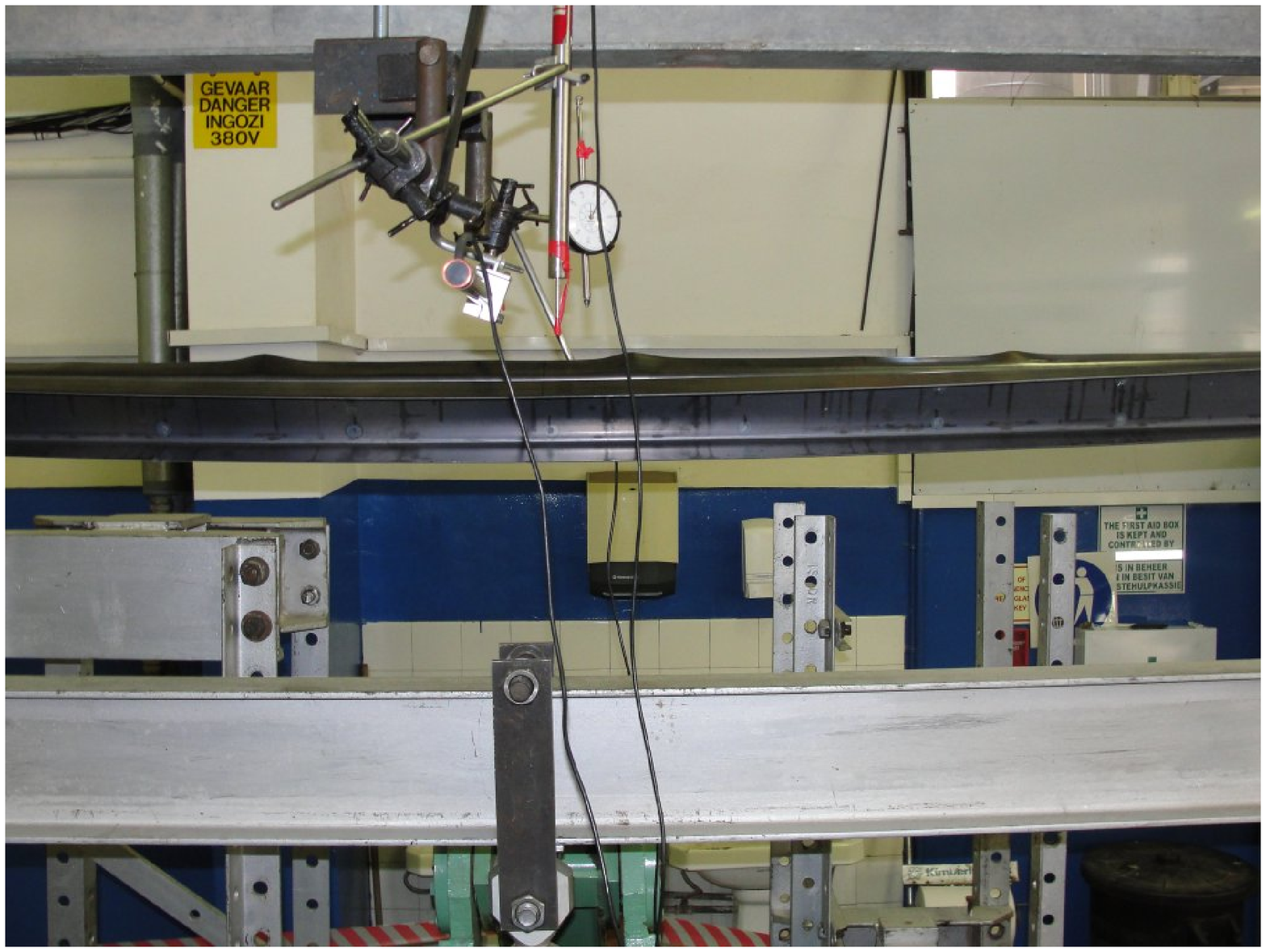,width=74mm}}
\subfigure[Test 6]{%
  \psfig{figure=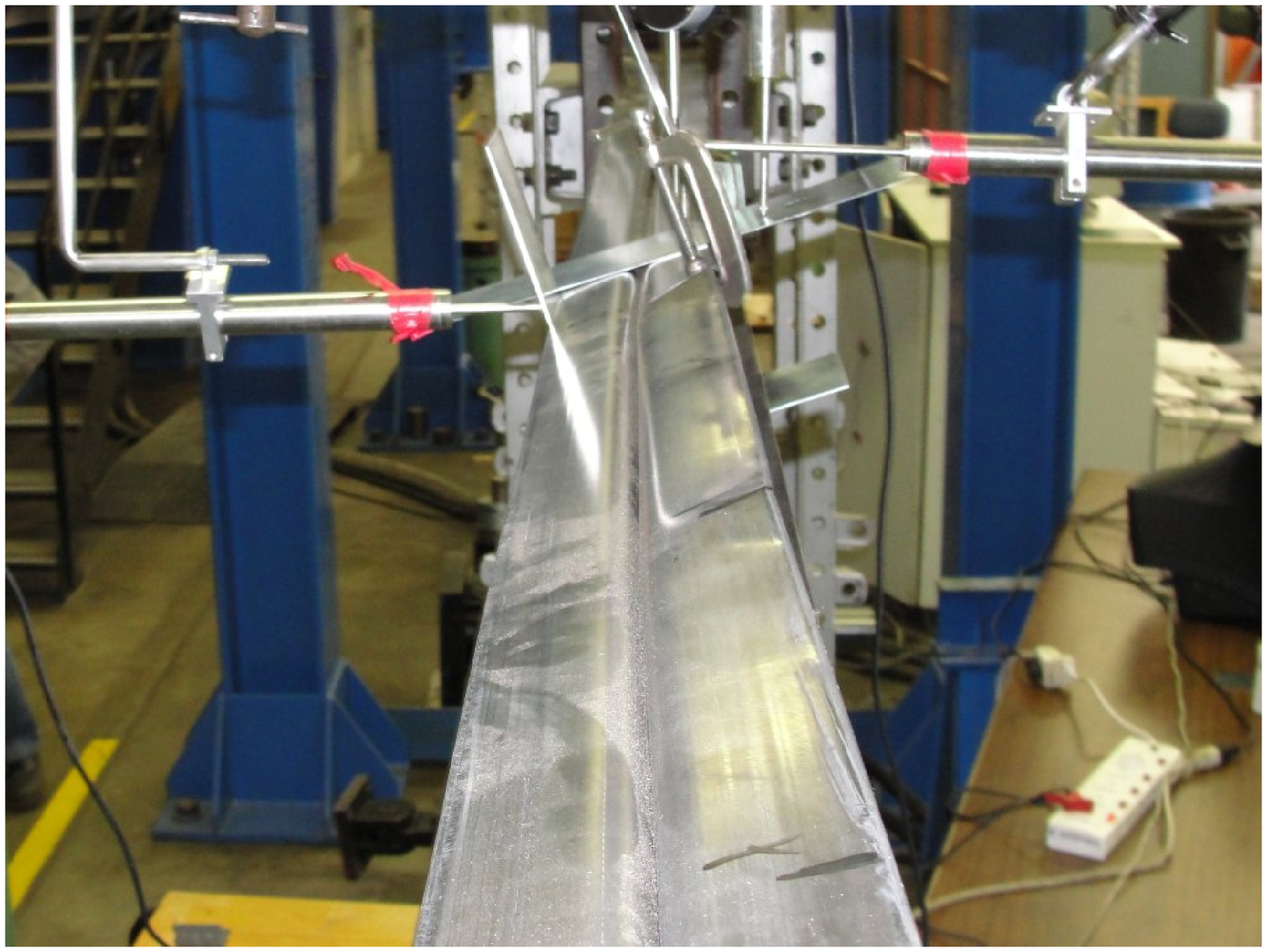,width=74mm}}
\subfigure[Post-testing beams: Tests 1, 3, 4 and 5]{%
  \psfig{figure=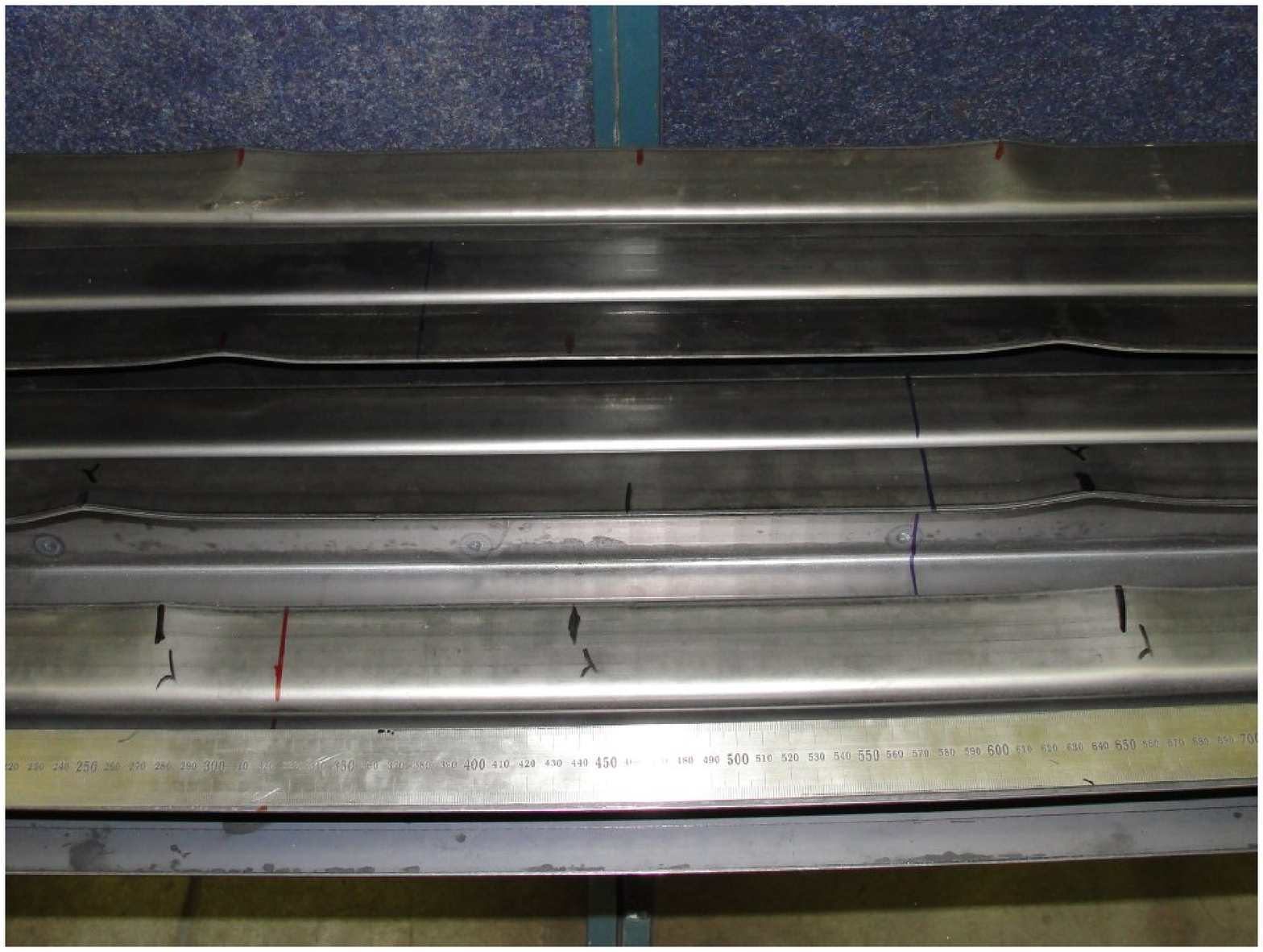,width=74mm}}
\caption{Selection of photographs from the experimental programme:
  (a)--(c) all showing interactive buckling and (d) shows four of the
  beams and their locally buckled flanges that show plastic
  deformation.}
\label{fig:exptphotos}
\end{figure}

\begin{figure}[htbp]
  \centering 
  \psfig{figure=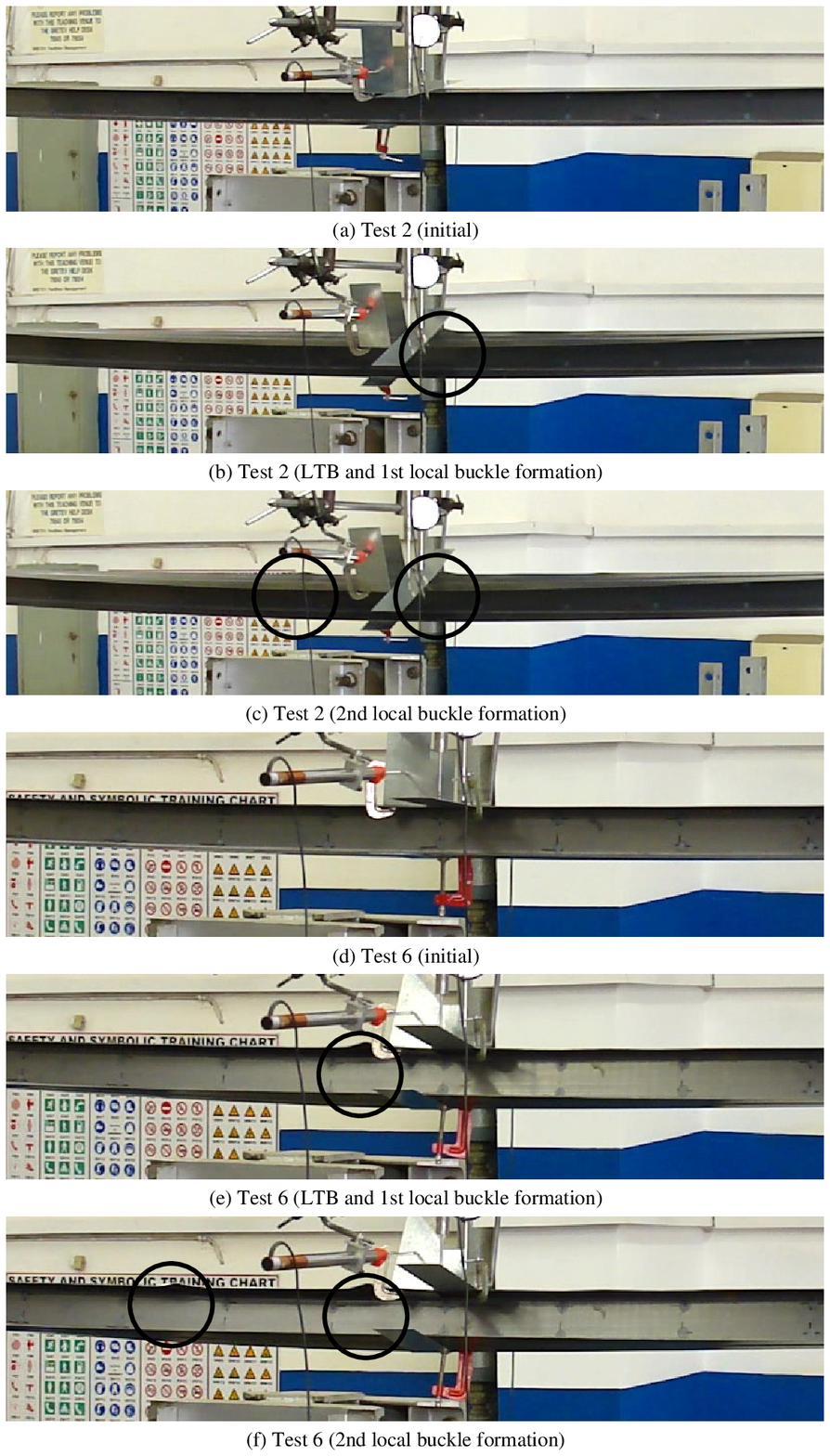,height=190mm}
  \caption{Evidence of cellular buckling. Two sequences of three
    photographs are shown of tests 2 (a)--(c) and 6 (d)--(f)
    respectively. Photographs (a) and (d) respectively show the
    pre-buckling state; (b) and (e) respectively show the initial
    post-buckling with one significant peak at midspan; (d) and (f)
    respectively show a newly developed local buckling peak in the
    top flange.}
  \label{fig:cellular}
\end{figure}

For each of the physical experiments performed in the present study,
see \S\ref{sec:expts}, testing proceeded to failure and all of the
beams exhibited an unstable response once interactive buckling was
triggered. A selection of photographs is presented in Figure
\ref{fig:exptphotos}
which show the beams from a variety of directions while they were
undergoing interactive buckling. In tests 2 and 6, there was visual
experimental evidence of cellular buckling; Figure \ref{fig:cellular}
shows a sequence of photographs before and after the principal
instability showing a new local buckling peak appearing soon after
the initial one. Table \ref{tab:Mmax_expt}
\begin{table}[htbp]
  \centering
  \begin{tabular}{c||c||c||c|c}
    & & Initial post-buckling &
    \multicolumn{2}{|c}{Local buckling mode observations} \\
    Test & $M_\mathrm{max} / M^\mathrm{C}$ & drop in moment (\%) &
    No. of visible & Extent \\
    & & & buckling peaks & (\% of $L_e$)\\
    \cline{1-5}
    $1$ & 1.03 & 15 & 5 & 25.4 \\
    $2$ & 0.90 & 7 & 7 & 36.5\\
    $3$ & 0.93 & 12 & 7 & 40.6\\
    $4$ & 1.07 & 24 & 8 & 52.4\\
    $5$ & 1.00 & 14 & 9 & 65.8\\
    $6$ & 1.05 & 15 & 9 & 69.2\\
  \end{tabular}
  \caption{Results from the experimental programme in terms of the
    maximum moment and the local buckling profile. The final column is
    a measure of localized nature of the flange buckle -- the smaller
    the number, the more it was localized.}
  \label{tab:Mmax_expt}
\end{table}
presents the results and their comparison with the individual buckling
modes. The maximum applied moment in the experiments $M_\mathrm{max}$
is presented as a ratio of the theoretical critical moment
$M^\mathrm{C}$ calculated from the appropriate critical mode given in
Table \ref{tab:Le}, whether LTB or local
buckling. The local buckling profile was determined by marking (as
seen in Figure \ref{fig:exptphotos}(d)) and measuring between adjacent
peaks of the local buckling displacement over the length of the
vulnerable part of the compression flange while the beam was still
loaded but well after the peak moment had been applied. The
interactive mode was clearly modulated in each case, with the peak
amplitudes from local buckling decaying towards the lateral
restraints; this was particularly notable in the cases where
LTB was critical since the number of peaks was
visibly fewer. In each test, two or three peaks of the local buckling
mode exhibited significant plastic deformation almost immediately
after the interactive mode was triggered; it was adjacent to these
peaks where the buckling wavelength was, in general, measured to be
the smallest values.

Another notable feature shown in Table \ref{tab:Mmax_expt} is the
immediate proportional drop in the moment once the interactive mode
had been triggered. As would be expected from the literature
\cite{IUTAM76_longer}, the largest drop occurred in test 4 where the
critical modes had been practically simultaneous. It is also
noteworthy that the tests with identical buckling lengths (tests 1 and
2) showed very different peak moments and moment drops. A rational
hypothesis can be devised for this by postulating that the beam in
test 2 contained more geometric imperfections than the beam in test
1. This would not only account for the smaller maximum moment measured
in test 2, but also for its smaller relative moment drop and its lower
residual moment in the post-buckling range \cite{TH73}.

\subsubsection{Comparisons with variational model and discussion}

\begin{figure}[htbp]
  \centering
  \subfigure[Tests 1 and 2: $L_e = 3200~\mm$]{%
    \psfig{figure=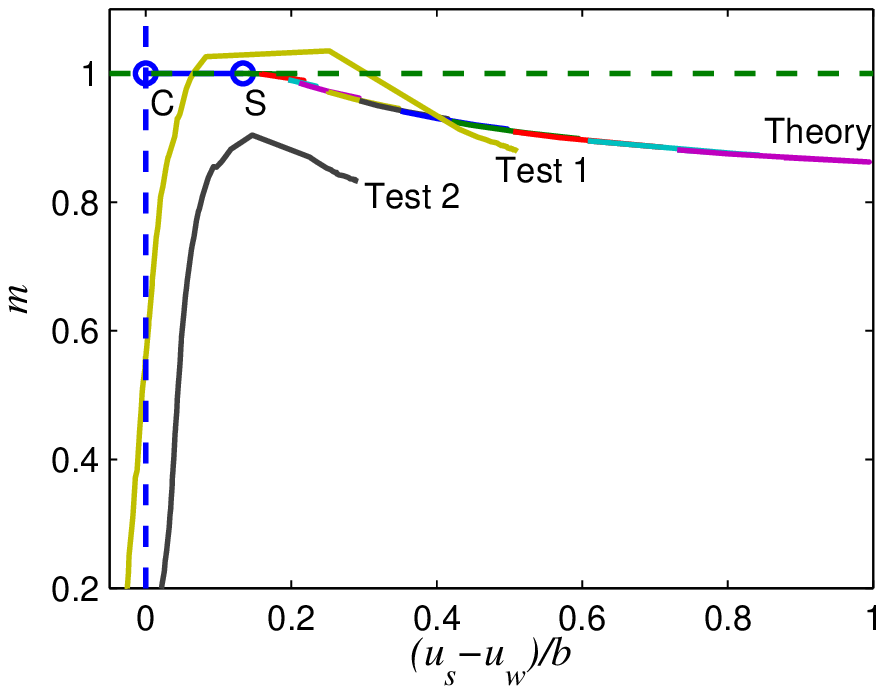,width=60mm}}\quad
  \subfigure[Test 3: $L_e = 3000~\mm$]{%
    \psfig{figure=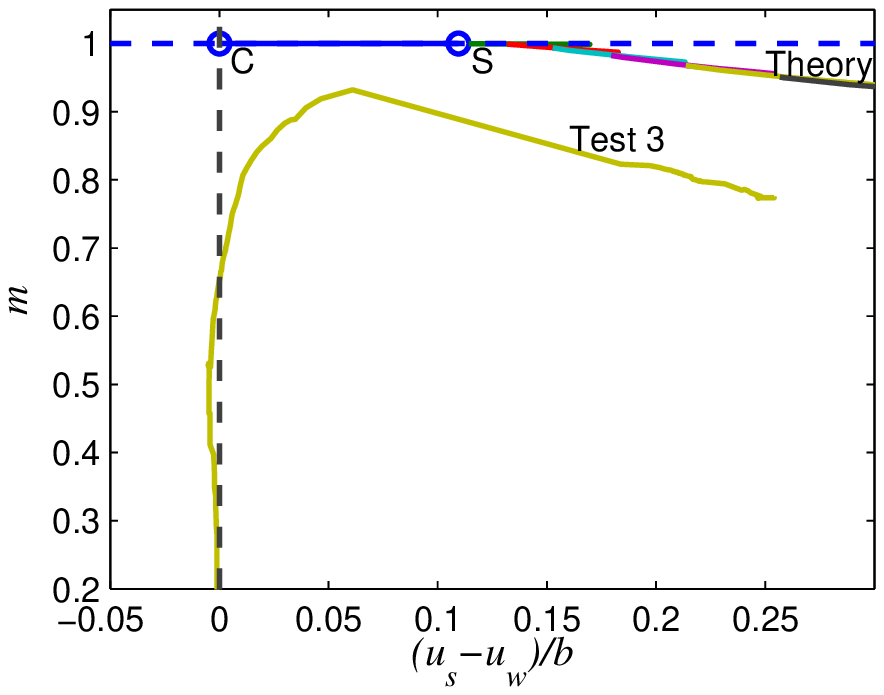,width=60mm}}\\
  \subfigure[Test 4: $L_e = 2750~\mm$]{%
    \psfig{figure=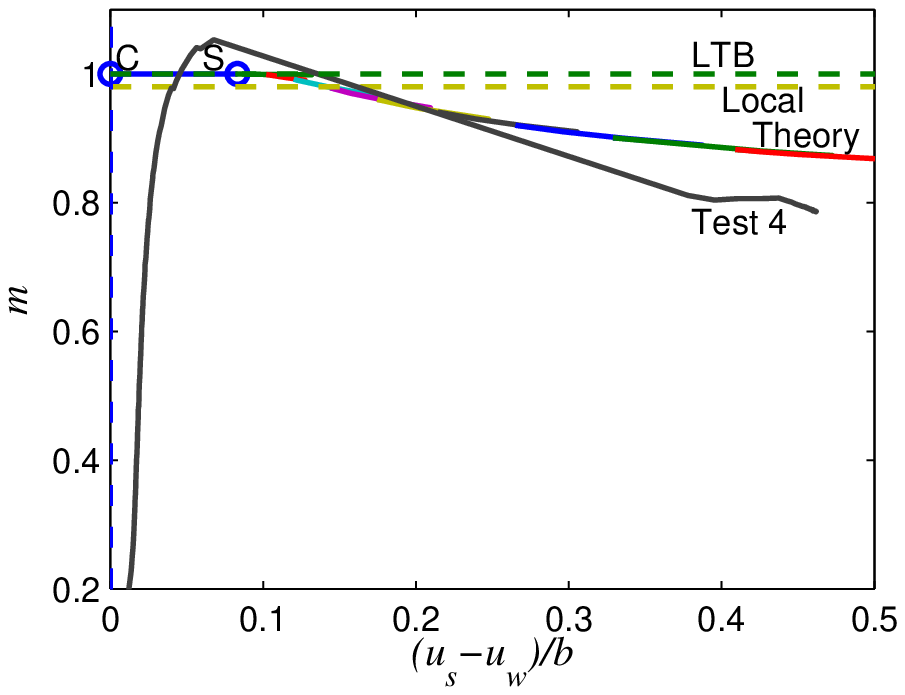,width=60mm}}\quad
  \subfigure[Test 5: $L_e = 2500~\mm$]{%
    \psfig{figure=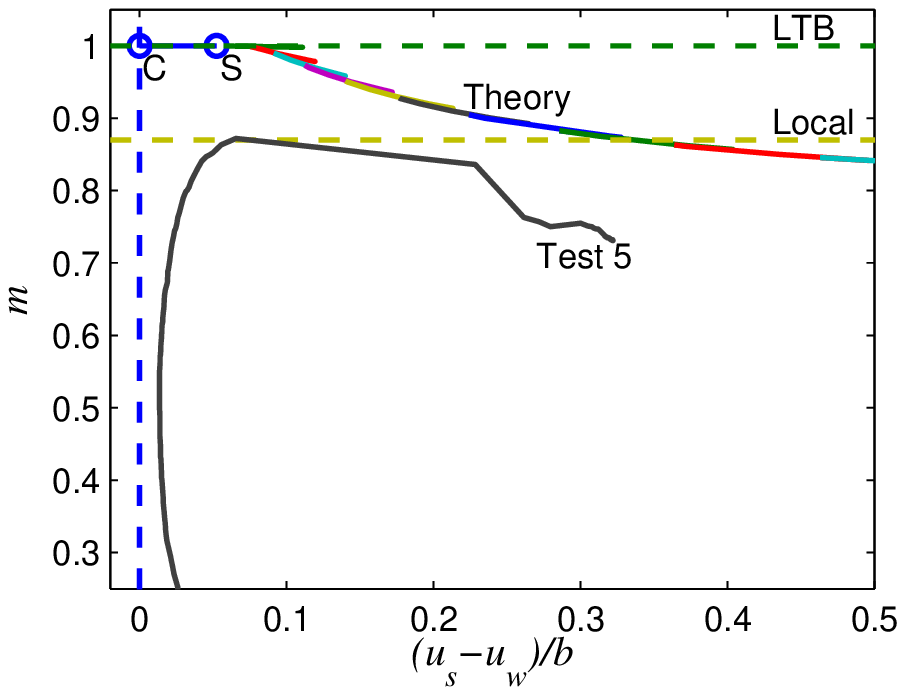,width=60mm}}\\
  \subfigure[Test 6: $L_e = 2250~\mm$]{%
    \psfig{figure=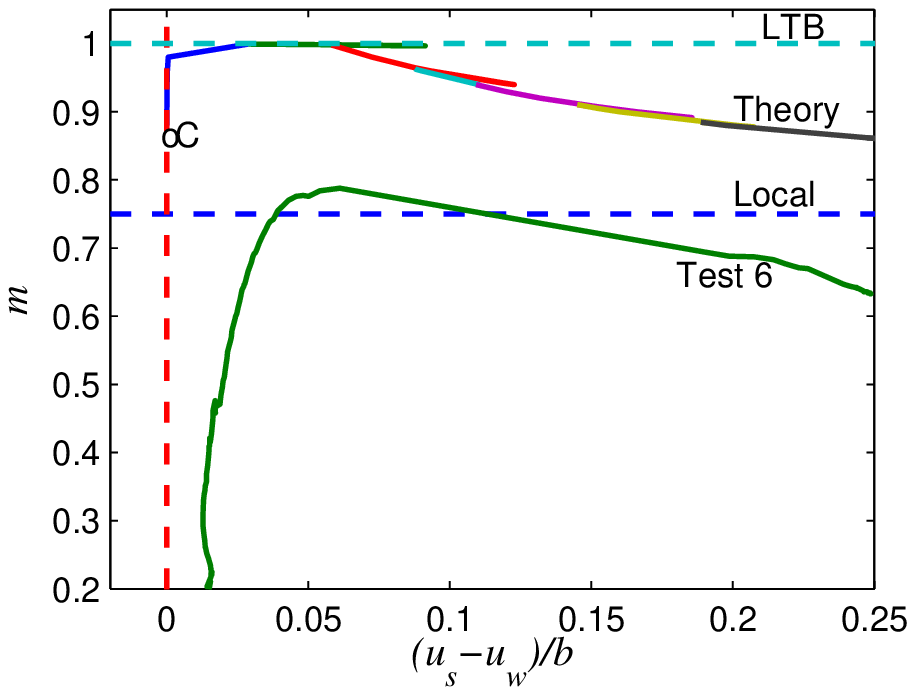,width=60mm}}
  \caption{Moment ratio $m$ versus bottom flange lateral displacement
    $(u_s-u_w)/b$ for all the experiments with the variational model
    predictions superimposed, denoted as ``Theory'' and evaluated by
    \textsc{Auto}. Points $\mathrm{C}$ and $\mathrm{S}$ refer to the
    critical and secondary instability points obtained from the model.
    Note that local buckling was the theoretical critical instability
    mode for Tests 4--6.}
  \label{fig:Mvsbot}
\end{figure}

Figure \ref{fig:Mvsbot}
presents normalized plots of the applied moment $m$ versus the
measured and normalized lateral displacement of the bottom flange,
$(u_s-u_w)/b$. Test 4 gives clearly the best comparison in terms of
the correlation between the post-buckling response of the actual beam
and the model prediction. Tests 1 and 2 also show good basic agreement
with the theory; test 1 showing that the post-buckling unloading
resembles the theory quite well, while test 2 shows that the
instability is triggered at a similar value of the lateral
displacement predicted by the theory (see Table \ref{tab:usuw}).
\begin{table}[htb]
  \centering
  \begin{tabular}{c|c|c||c|c|c}
    Test & \multicolumn{2}{|c||}{$(u_s-u_w)/b$} &
    \multicolumn{3}{|c}{Local buckling wavelengths $\Lambda$ ($\mm$)} \\
    & Expt & Theory & Expt range & Expt average
    & Theory (minimum)\\
    \cline{1-6}
    $1$ & 0.169 & 0.134 & 185 $\rightarrow$ 221 & 203 & 200\\
    $2$ & 0.146 & 0.134 & 171 $\rightarrow$ 250 & 195 & 200\\
    $3$ & 0.061 & 0.110 & 161 $\rightarrow$ 242 & 203 & 150\\
    $4$ & 0.068 & 0.083 & 161 $\rightarrow$ 249 & 206 & 153\\
    $5$ & 0.066 & 0.052 & 167 $\rightarrow$ 247 & 206 & 163\\
    $6$ & 0.061 & 0.058 & 158 $\rightarrow$ 225 & 195 & 240\\
  \end{tabular}
  \caption{Comparisons of the experimental results with the
    variational model in terms of the values of the bottom flange
    displacement at the peak moment point and the local buckling
    wavelengths.}
  \label{tab:usuw}
\end{table}
Tests 5 and 6 clearly peak at or marginally above the local buckling
critical moment, as predicted from linear analysis. However, in a
similar way to test 2, the instability is triggered at a lateral
displacement that correlates well with the prediction from the
variational model. For test 6, the variational model yields a lower
critical moment than the $\Mcr$ value for LTB, which triggers a
quasi-local buckling mode. However, as stated earlier, a distinct and
accurate local buckling mode can only be modelled with additional
displacement functions in the current framework so this particular
result needs to be interpreted with some caution. Test 3 could be
considered to be an outlier, but the measured response would imply, in
a similar way that was discussed above regarding test 2, that the
level of geometric imperfections in this beam was higher than the
other tested beams (1, 4, 5, 6). Hence, the measured instability
moment is less, the unloading proportion is less and the response is
practically parallel to the model curve, which in fact is encouraging.

In terms of the local buckling wavelengths, these are compared to the
wavelength of the buckling profile obtained from the variational model
as described in \S\ref{sec:existing}.  Even though the theoretical
results seemed to be influenced by effects close to the boundary
(particularly in test 6), hence the variability in the predictions,
the general correlation between the experiments and theory is good.
The apparent confirmation that the post-buckling behaviour of an
I-beam under pure bending is cellular when global and local
instability modes interact nonlinearly poses the following question:
is this phenomenon prevalent in other thin-walled structural
components that are known to suffer from overall and local mode
interaction? Compressed stringer-stiffened plates \cite{Koiter76} and
I-section struts \cite{Becque_Rasmussen_expt_ASCE2009} are prime
examples of other components where local and global mode interactions
are known to occur. Further research is obviously required to
determine the answer.

\section{Concluding remarks}

The current work identifies an interactive form of buckling for an
I-beam under uniform bending which couples a global instability with
local buckling in one-half of the compression flange. In contrast to
earlier, more numerical, work \cite{Mollmann2_ijss_89,MSGP97},
cellular buckling, the transformation from a localized to an
effectively periodic mode, is predicted theoretically for the purely
elastic case and evidenced in physical tests.  The model compares well
both qualitatively and quantitatively with the observed collapse of a
beam that undergoes the interaction under discussion that involves
global, local, localized and cellular buckling. The localized buckle
pattern first appears at a secondary bifurcation point which
immediately destabilizes a portion of the compression flange; as the
deformation grows, the buckle tends to spread in cells until
eventually it restabilizes when the localized buckling pattern has
become periodic after a sequence of snap-backs.

Experimentally, the process is unstable and so this sequence occurs
rapidly even under rigid loading with the local buckling cells being
triggered dynamically. This highlights the practical dangers of the
modelled and observed phenomenon; the interaction reduces the load
carrying capacity, it therefore introduces an imperfection sensitivity
that would need to be quantified such that robust design rules can be
developed to mitigate against such hazardous structural behaviour.

\subsubsection*{Acknowledgements}

The majority of this work was conducted while MAW was
on sabbatical from April--October 2010 at the School of Civil and
Environmental Engineering, University of the Witwatersrand,
Johannesburg, South Africa. The authors are extremely grateful to the
Head of School, Professor Mitchell Gohnert, the Senior Laboratory
Technician, Kenneth Harman, and Spencer Erling from the South African
Institute of Steel Construction for facilitating the experimental
programme.

\bibliography{allrefs}

\end{document}